\documentstyle[aps,prb,multicol]{revtex}

\input {epsf}

\newcommand{\eqref}[1]{(\ref{#1})}

\newcommand{\half}{\frac{1}{2}}

\renewcommand{\Re}{\mbox{Re}\,}

\newcommand{\sign}{\mbox{sign}}
\newcommand{\plus}{\; \; \, }
\newcommand{\mfp}{\mbox{$\ell $}}

\newcommand{\p}{{\bf p}}
\newcommand{\pf}{{\bf p}_f}
\newcommand{\vf}{{\bf v}_f}

\newcommand{\R}{{\bf R}}

\renewcommand{\vec}[1]{{\bf{#1}}}

\newcommand{\beq} {\begin{equation}}
\newcommand{\eeq} {\end{equation}}
\newcommand{\ber} {\begin{eqnarray}}
\newcommand{\eer} {\end{eqnarray}}
\newcommand{\nn}{\nonumber }

\newcommand{\de}   {{\delta}}

\newcommand{\epp}[2] {\mbox{$#1\!+\!{#2\over 2}$}}
\newcommand{\epm}[2] {\mbox{$#1\!-\!{#2\over 2}$}}

\newcommand{\vars} {\pf,\R,\ep,t}
\newcommand{\name}[1] {{#1}}

\newcommand{\tc}{\op{\tau }_3 }

\newcommand{\tz}{\op{0} }

\newcommand{\ep}{\epsilon }
\newcommand{\om}{\omega }

\newcommand{\mat}{\left(\begin{array}{cc}}
\newcommand{\matend}{\end{array}\right)}
\newcommand{\matl}{\left(\begin{array}{c}}
\newcommand{\matlend}{\end{array}\right)}
\newcommand{\qt}{{\, \scriptstyle \otimes}\, }

\newcommand{\op}[1]{\hat{#1}}

\newcommand{\PPp}{\check{P}_{+}}
\newcommand{\PPm}{\check{P}_{-}}
\newcommand{\PPpm}{\check{P}_{\pm}}

\newcommand{\Pa}{\op{P}_{+}}
\newcommand{\Pb}{\op{P}_{-}}
\newcommand{\Par}{\Pa^{\ret}}
\newcommand{\Pbr}{\Pb^{\ret}}
\newcommand{\Paa}{\Pa^{\adv}}
\newcommand{\Pba}{\Pb^{\adv}}
\newcommand{\Para}{\Pa^{\ra}}
\newcommand{\Pbra}{\Pb^{\ra}}

\newcommand{{\ra}}{{\scriptscriptstyle R,A}}
\newcommand{{\rak}}{{\scriptscriptstyle R,A,K}}
\newcommand{{\ret}}{{\scriptscriptstyle R}}
\newcommand{{\adv}}{{\scriptscriptstyle A}}
\newcommand{\kel}{{\scriptscriptstyle K}}

\newcommand{\ano}{{\scriptstyle a}}

\newcommand{\gq}{\op{g}}
\newcommand{\gqr}{\gq^{\ret}}
\newcommand{\gqa}{\gq^{\adv}}
\newcommand{\gqra}{\gq^{\ra}}
\newcommand{\gqk}{\gq^{\kel}}

\newcommand{\dgqr}{\de \! \gqr}
\newcommand{\dgqa}{\de \! \gqa}
\newcommand{\dgqra}{\de \! \gqra}
\newcommand{\dgqk}{\de \! \gqk}
\newcommand{\dgqan}{\de \! \gq^{\ano}}
\newcommand{\gaq}{g}
\newcommand{\gbq}{\tilde{g}}

\newcommand{\gaqra}{\gaq^{\ra}}
\newcommand{\gbqra}{\gbq^{\ra}}
\newcommand{\gaqk}{\gaq^{\kel}}
\newcommand{\gbqk}{\gbq^{\kel}}

\newcommand{\faq}{f}
\newcommand{\fbq}{\tilde{f}}

\newcommand{\faqra}{\faq^{\ra}}
\newcommand{\fbqra}{\fbq^{\ra}}
\newcommand{\faqk}{\faq^{\kel}}
\newcommand{\fbqk}{\fbq^{\kel}}

\newcommand{\ppa}{\rho}
\newcommand{\ppb}{\tilde\ppa}
\newcommand{\ppar}{\ppa^{\ret}}

\newcommand{\ppbr}{\ppb^{\ret}}

\newcommand{\ga}{\gamma}
\newcommand{\gb}{\tilde{\ga}}
\newcommand{\gar}{\ga^{\ret}}
\newcommand{\gbr}{\gb^{\ret}}
\newcommand{\gaa}{\ga^{\adv}}
\newcommand{\gba}{\gb^{\adv}}
\newcommand{\gara}{\ga^{\ra}}
\newcommand{\gbra}{\gb^{\ra}}
\newcommand{\Ga}{\Gamma}
\newcommand{\Gb}{\tilde{\Ga}}
\newcommand{\Gar}{\Ga^{\ret}}
\newcommand{\Gbr}{\Gb^{\ret}}
\newcommand{\Gaa}{\Ga^{\adv}}
\newcommand{\Gba}{\Gb^{\adv}}
\newcommand{\Gara}{\Ga^{\ra}}
\newcommand{\Gbra}{\Gb^{\ra}}

\newcommand{\dgar}{\de \! \gar}
\newcommand{\dgbr}{\de \! \gbr}
\newcommand{\dgaa}{\de \! \gaa}
\newcommand{\dgba}{\de \! \gba}
\newcommand{\dgara}{\de \! \gara}
\newcommand{\dgbra}{\de \! \gbra}

\newcommand{\dGar}{\de \! \Gar}
\newcommand{\dGbr}{\de \! \Gbr}
\newcommand{\dGaa}{\de \! \Gaa}
\newcommand{\dGba}{\de \! \Gba}

\newcommand{\xa}{x^{\kel}}
\newcommand{\xb}{\tilde{x}^{\kel}}
\newcommand{\dxak}{\de \! x^{\kel}}
\newcommand{\dxaan}{\de \! x^{\ano}}
\newcommand{\dxban}{\de \! \tilde{x}^{\ano}}
\newcommand{\dxbk}{\de \! \tilde{x}^{\kel}}
\newcommand{\dXak}{\de \! X^{\kel}}
\newcommand{\dXbk}{\de \! \tilde{X}^{\kel}}
\newcommand{\Xa}{X^{\kel}}
\newcommand{\Xb}{\tilde{X}^{\kel}}
\newcommand{\Fe}{\Fa_{eq}}
\newcommand{\Fa}{F}
\newcommand{\Fb}{\tilde{\Fa}}
\newcommand{\dxa}{\de \! x^{\ano}}
\newcommand{\dxb}{\de \! \tilde{x}^{\ano}}
\newcommand{\dXa}{\de \! X^{\ano}}
\newcommand{\dXb}{\de \! \tilde{X}^{\ano}}

\newcommand{\XX}{\op{X}^{\kel}}
\newcommand{\dXXra}{\de \! \op{X}^{\ra}}
\newcommand{\dYYra}{\de \! \op{Y}^{\ra}}
\newcommand{\dXXan}{\de \! \op{X}^{\ano}}
\newcommand{\dYYan}{\de \! \op{Y}^{\ano}}
\newcommand{\YY}{\op{Y}^{\kel}}

\newcommand{\Nr}{\op{N}^{\ret}}
\newcommand{\Na}{\op{N}^{\adv}}
\newcommand{\Nra}{\op{N}^{\ra}}

\newcommand{\Da}{\Delta}
\newcommand{\Dm}{\op{\Da }}
\newcommand{\Db}{\tilde{\Da}}
\newcommand{\Dar}{\Da^{\ret}}
\newcommand{\Dbr}{\Db^{\ret}}
\newcommand{\Daa}{\Da^{\adv}}
\newcommand{\Dba}{\Db^{\adv}}
\newcommand{\Dara}{\Da^{\ra}}
\newcommand{\Dbra}{\Db^{\ra}}

\newcommand{\Dak}{\Da^{\kel}}
\newcommand{\Dbk}{\Db^{\kel}}

\newcommand{\dDara}{\de \! \Dara}
\newcommand{\dDbra}{\de \! \Dbra}

\newcommand{\dDaan}{\de \! \Da^{\ano}}
\newcommand{\dDban}{\de \! \Db^{\ano}}

\newcommand{\sma}{\sigma}

\newcommand{\va}{\Sigma}
\newcommand{\vb}{\tilde{\va}}
\newcommand{\vak}{\va^{\kel}}
\newcommand{\vbk}{\vb^{\kel}}
\newcommand{\vara}{\va^{\ra}}
\newcommand{\vbra}{\vb^{\ra}}
\newcommand{\var}{\va^{\ret}}
\newcommand{\vbr}{\vb^{\ret}}
\newcommand{\vaa}{\va^{\adv}}
\newcommand{\vba}{\vb^{\adv}}

\newcommand{\dvara}{\de \! \vara}
\newcommand{\dvbra}{\de \! \vbra}

\newcommand{\dvaan}{\de \! \va^{\ano}}
\newcommand{\dvban}{\de \! \vb^{\ano}}

\newcommand{\Ea}{\varepsilon}

\newcommand{\Eara}{\Ea^{\ra}}

\newcommand{\hk}{\op{h}^{\kel}}
\newcommand{\hr}{\op{h}^{\ret}}
\newcommand{\ha}{\op{h}^{\adv}}
\newcommand{\hra}{\op{h}^{\ra}}
\newcommand{\gmk}{\op{g}^{\kel}}
\newcommand{\gmr}{\op{g}^{\ret}}
\newcommand{\gma}{\op{g}^{\adv}}
\newcommand{\gmra}{\op{g}^{\ra}}
\newcommand{\smm}{\op{\sma}}
\newcommand{\smmk}{\op{\sma}^{\kel}}
\newcommand{\smmr}{\op{\sma}^{\ret}}
\newcommand{\smma}{\op{\sma}^{\adv}}

\newcommand{\dhk}{\de \! \hk}
\newcommand{\dhan}{\de \! \op{h}^{\ano}}
\newcommand{\dhr}{\de \! \hr}
\newcommand{\dha}{\de \! \ha}
\newcommand{\dhra}{\de \! \hra}

\newcommand{\qpartial}{\vf \! \mbox{\boldmath $\nabla $}}

\newcommand{\ce}{\check{1}}
\newcommand{\cz}{\check{0}}
\newcommand{\cg}{\check{g}}

\renewcommand{\Re}{\mbox{Re}}
\newcommand{\lrule}{ \end{multicols} \noindent
  \rule{0.5\textwidth}{0.1mm}\rule{0.1mm}{3pt}\newline }
\newcommand{\rrule}{ \noindent \parbox{\textwidth}{
  \hfill\rule[-3pt]{0.1mm}{3pt}\rule{0.5\textwidth}{0.1mm}}
  \begin{multicols}{2} }
\newcommand{\state}{coherence }
\newcommand{\State}{COHERENCE }

\begin{document}

\draft
\title{Distribution functions in non-equilibrium theory of superconductivity\\
and Andreev spectroscopy in unconventional superconductors }
\author{Matthias Eschrig}
\address{
Department of Physics \& Astronomy, Northwestern University, Evanston, IL 60208, USA}
\date{\today}
\maketitle
\begin{abstract}
\noindent
We present a new theoretical formulation of  non-equilibrium  
superconducting phenomena, including singlet and triplet pairing.
We start from the general  
Keldysh-Nambu-Gor'kov Green's functions in the quasiclassical
approximation and represent them in terms of 
2x2 spin-matrix \state functions and distribution functions for particle-type 
and hole-type excitations. The resulting transport equations for the 
distribution functions may be  interpreted as a generalization to the
superconducting state of Landau's
transport equation for the normal Fermi liquid of  conduction electrons.
The equations  are 
well suited for numerical simulations of dynamical phenomena.
Using our formulation we solve an open problem in quasiclassical theory
of superconductivity, the derivation of an explicit representation of
Zaitsev's nonlinear boundary conditions\cite{zaitsev84} 
at surfaces and interfaces. These boundary conditions include non-equilibrium
phenomena and spin singlet and triplet unconventional pairing.
We eliminate spurious  solutions as well as numerical stability problems 
present in the original formulation. 
Finally, we formulate the Andreev scattering problem at 
interfaces in terms of the introduced distribution functions and
present a theoretical analysis for the study of time reversal symmetry 
breaking states in unconventional superconductors via
Andreev spectroscopy experiments at N-S interfaces with finite transmission.
We include impurity scattering self consistently.
\end{abstract}
\pacs{PACS numbers:  74.20.-z, 74.80.Fp, 74.50.+r}
\vspace{-11pt}
\begin{multicols}{2}
\section{Introduction}\label{sect_introduction}

\vspace{-11pt}
Conduction electrons in normal metals 
are  generally well 
described by Landau's Fermi liquid theory.\cite{landau57}
 According to Landau  a system  of
strongly interacting electrons 
may be viewed as  
an ensemble  of quasiparticles 
which   can be 
represented by a  classical distribution function and obeys   
a semiclassical Landau-Boltzmann transport 
equation.\cite{landau57}
This  semiclassical 
behavior of a quantum many-body system is a consequence of
 Pauli's  exclusion principle which restricts the 
momentum space accessible to    low-energy quasiparticles to
a thin shell near the Fermi surface. The ratio of the volume of the 
accessible momentum 
space to the total volume enclosed by  the  Fermi surface is of the order 
$k_BT/E_f\ll 1$, and is    the   fundamental   expansion 
parameter of Fermi liquid theory.
Landau's Fermi liquid theory 
is exact  in leading order in an  asymptotic expansion in $k_BT/E_F$
and other
  small parameters of an  electronic Fermi liquid such as
 $1/k_f\xi_T$, $\hbar \omega/E_f$, $1/k_f\ell$, 
where $E_f$, $k_f$, $\xi_T$, $\omega$, and $\ell$  are Fermi energy, Fermi
wave vector, thermal coherence length ($\xi_T=\hbar v_f/2\pi k_BT$), frequency 
of time-dependent perturbations, and quasiparticle mean free path.
Phase space arguments can be used to derive Landau's Fermi liquid theory 
by converting  a formal 
diagrammatic expansion
of many-body Green's functions into 
 an asymptotic  expansion 
in the above small parameters.\cite{Eliashberg62,eliashberg71,serene83}
Only a few of the resummed Feynman self-energy diagrams 
contribute in  leading orders, and the dynamical equations 
 for Green's functions can 
be transformed into Landau's transport equation 
for quasiparticle distribution
functions.\cite{Abrikosov59,Landau59,Eliashberg62,eliashberg71} 
The price one has to pay for the simplifications of the quasiparticle theory 
is the need to introduce  
phenomenological parameters, such as the quasiparticle velocities 
and quasiparticle interactions.\cite{landau57,Landau59} In the absence of first
principles  calculations
these material parameters  have
to be taken from the experiment.

The development of a generalized Fermi-liquid theory 
for the superconducting state  started 
with the work by 
\name{Geilikman}\cite{Geilikman58,Geilikman58a}
and \name{Bardeen} {\em et al.}\cite{Bardeen59}
shortly after 
the BCS-theory of superconductivity was published.\cite{bardeen57}
These authors presented a generalization of the 
semiclassical transport equations of the normal state
to the  superconducting state and used them to explain the
electronic thermal conductivity of superconductors.
Several early  works\cite{Larkin63,Ambegaokar64,Leggett65} on 
transport and linear response  in superconductors
showed that various  semiclassical concepts of Landau's  Fermi liquid
theory could be readily generalized to  the superconducting 
state. A novel  semiclassical 
approach to superconductors in equilibrium 
was initiated  by \name{de Gennes}\cite{degennes66}
who formulated the 
equilibrium theory of superconductivity near T$_c$ in terms of classical
correlation functions. These developments  established the usefulness and accuracy of
semiclassical concepts for superconductors but did not provide a complete semiclassical
theory of superconductivity.
Nevertheless,  these early semiclassical works   were the predecessors of
a comprehensive  theory developed 
by \name{Eilenberger},\cite{eilenberger68} and
\name{Larkin} and \name{Ovchinnikov}\cite{larkin68}
for superconductors in equilibrium.
This theory was  generalized to 
non-equilibrium phenomena  by \name{Eliashberg}\cite{eliashberg71} and  
\name{Larkin} \& \name{Ovchinnikov}.\cite{larkin75} We 
 regard this theory as
the proper generalization of Landau's Fermi liquid theory to the
superconducting state, and call this theory, 
following \name{Larkin} and \name{Ovchinnikov}\cite{larkin68}, the
{\em quasiclassical theory of superconductivity}. 

 The derivation 
of the quasiclassical equations starts from
 Gor'kov's  formulation of the theory of superconductivity
in terms of Nambu-Gor'kov matrix Green's functions.\cite{gorkov58}
Typical spatial variations of the order parameter occur on a scale of the
coherence length, $\xi_0=\hbar v_f/2\pi k_BT_c$,
and  the typical time scale, $t_0$, is given by the
inverse gap, $t_0\sim \hbar/\Delta $. Both scales are usually  large
in superconducting metals as compared to the corresponding atomic 
scales, $\sim 1/k_f$, and $\sim \hbar/E_f$.
Gor'kov's Green's functions contain detailed information on 
atomic scale properties which average out on the superconducting scales.
To derive  the quasiclassical equations one has  to integrate out 
atomic scale features in the Green's functions, but keep all relevant 
information  for superconductivity.
The resulting ``$\xi$-integrated''  Green's functions 
(quasiclassical Green's functions) vary on
the large scales $\xi_0 $ and $t_0$ and are free of the irrelevant fine-scale
structures. The quasiclassical equations should be compared
with Andreev's equations\cite{andreev64} which he obtained by
factorizing  out  rapidly  oscillating
terms in  \name{Bogoljubov's} equations.\cite{Bogoliubov58,degennes66} 
 \name{Andreev's} method is  equivalent to the
 quasiclassical theory for   superconductors 
with infinitely long-lived quasiparticles, i.e. without 
impurities, electron-phonon coupling or electron-electron scattering. 
Both theories give identical results in these cases.
 Hence, the quasiclassical theory of 
\name{Eilenberger}, \name{Larkin}, 
\name{Ovchinnikov}, and \name{Eliashberg}
may  be considered a generalization of 
\name{Andreev's} equations to systems with  disorder and finite lifetimes of
quasiparticles. The generalized theory covers
basically all phenomena of interest in superconductivity. 

One distinguishes  in the quasiclassical theory of superconductivity 
external (classical) and internal
(quantum mechanical) degrees of freedom of a quasiparticle. Quasiparticles move 
along classical trajectories, and
are reflected by walls or other obstacles and
scattered by collisions with impurities, phonons or with other
quasiparticles. This classical  dynamics in coordinate space 
should be contrasted with the
quantum-dynamics of  internal degrees of freedom of a quasiparticle which are the 
spin (S=$1\over 2$) and the particle-hole degree of freedom. The quantum coherence
of particle and hole excitations is an essential feature of BCS pairing theory.
It is absent in the normal state and is the origin of all typical
superconducting phenomena 
including the opening of an energy gap,\cite{bardeen57} Andreev's
retro-reflection,\cite{andreev64} Tomasch oscillations,\cite{Tomasch66}
vortex bound states,\cite{Caroli64} etc. 

In a standard notation of
quasiclassical theory\cite{serene83,rammer86,Larkin86} the distribution functions for quasiparticles 
are $4\times4$ matrices which reflects their quantum mechanical structure
as density matrices in the  4-dimensional Hilbert space of internal degrees of freedom
(Nambu-Gor'kov space).
The matrix elements are functions of the classical variables
${\bf p}_f$ (Fermi momentum), ${\bf R}$ (position), $\epsilon$ (energy measured from the
chemical potential), and $t$ (time),
which describes the classical degrees of freedom. The external motion of a quasiparticle
and its internal particle-hole state are coupled in a subtle way as first discussed by
Andreev.\cite{andreev64}

The complexity of dynamical phenomena in superconductors makes the elimination
of atomic scale fine structure an important step towards a solution of dynamical problems.
The dynamical equations of the quasiclassical theory, which are free of 
microscopic fine
structures from the outset, can be formulated most
compactly by using the Green's function technique of Keldysh.\cite{keldysh64}
It is often useful to distinguish
two sources of time-dependent phenomena. Firstly, time dependences can arise from changes in
the
occupation of quasiparticle states. In the normal phase of a Fermi liquid this is indeed the only
fundamental dynamical  process. Here, the quasiparticle states are robust and changes of the
quasiparticle wave function can be neglected. This is no longer the case in the
superconducting phase. Quasiparticle states in superconductors 
are coherent mixtures of particle and hole
states determined by the superconducting order parameter.
Since the order parameter will, in general,  change in a dynamical process
the quasiparticle states will also change. Superconducting
dynamics is thus governed by the coupled dynamics of both the quasiparticle states and
their occupation. 
The Keldysh technique is convenient in this case since it 
works with two types of Green's functions 
($g^{R,A}$ and $g^K$) and can be used to introduce dynamical spectral functions describing the
time development of quasiparticle states 
and dynamical distribution functions describing the
time-dependent occupation of the states.
Dynamical distribution functions in the superconducting 
state were introduced by 
Larkin \& Ovchinnikov\cite{larkin75} and by Shelankov.\cite{shelankov85}

In this paper we present an exact parameterization of the quasiclassical Keldysh Green's 
functions in terms of four \state functions and two distribution functions.
The \state functions are generalizations of the Riccati amplitudes introduced recently\cite{nagato93,schopohl95}
for superconductors in equilibrium, whereas the distribution functions
are the  generalizations of the distribution function of 
Landau's Fermi liquid theory of the normal state.
Compared to the conventional quasiclassical theory
our formulation leads to intuitively appealing 
and {\it explicit} boundary conditions at surfaces and 
interfaces, is numerically very stable, and allows for a more transparent
interpretation of quasiclassical dynamics in terms of particle-type and hole-type excitations.

The general framework of the
quasiclassical theory is briefly reviewed in sections II, III, where we also introduce our
notation.
Dynamical equations for the \state functions are derived in section IV together with
dynamical equations for the distribution functions (transport equations).
In section V we solve Zaitsev's non-linear boundary conditions for quasiclassical
Green's functions at interfaces, and obtain physically appealing boundary conditions
for our \state functions and distribution functions. 
In section VI we present 
the general linear response equations in terms of the introduced functions. 
Finally, we formulate in section VII the Andreev scattering problem at
interfaces between a normal metal and an unconventional superconductor
using the new theoretical formulation and the resulting boundary
conditions.
We present results for the Andreev reflection amplitudes and the
regularly reflected amplitudes at (110) interfaces and
(100) interfaces between a normal metal and a 
layered $d$-wave superconductor. 
Our formulation generalizes earlier 
work\cite{blonder82,bruder90} to include disorder.
We propose an anomalous feature in the reflection amplitudes for
(110) interfaces as a test for time reversal symmetry breaking states.
This feature, a strong suppression of the regular reflection 
for low-energy quasiparticles at interfaces with finite transmission, 
is sensitive to sign changes in the order parameter, and
has the same origin as the zero-energy surface bound states.
Combined with this suppression is an
enhancement of the excess current due to Andreev reflection for
low energy quasiparticles.
The sensitivity of this phenomenon to time reversal symmetry breaking
states provides a new tool to study the symmetry of the order parameter.
We study the effect of disorder on both
regularly and Andreev reflected currents.
For Andreev spectroscopy in unconventional superconductors the 
low-energy behavior of regular reflection is the 
spectral feature most stable against disorder.

\section{Keldysh space structure}

The fundamental quantity in non-equilibrium
quasiclassical theory of superconductivity is the quasiclassical
Green's function $\check{g}=\check{g}(\vars )$.\cite{serene83,rammer86,Larkin86}
It is a 2x2 Keldysh matrix\cite{keldysh64} of the form
\beq
\check{g}=\left( \begin{array}{cc} \gmr
& \gmk \\ \tz & \gma
\end{array} \right) ,\;
\eeq
where the elements are 4x4-Nambu matrices, which describe the two 
important residual quantum mechanical degrees of freedom: the
spin degree of freedom and the particle-hole degree of freedom.
$ \hat{g}^R=\hat{g}^R(\vars)$  is the retarded, $\hat{g}^A=\hat{g}^A(\vars)
$ the advanced and $\hat{g}^K=\hat{g}^K(\vars)$
the Keldysh Green's function. The classical degrees of freedom are described
by a motion of the quasiparticles along classical trajectories.
All trajectories through a spatial point $\vec{R} $ are parameterized by
the Fermi momentum, $\vec{p}_f$, and 
their directions coincide with the directions of the Fermi velocities,
$\vf (\vec{p}_f)$. Along a given trajectory with fixed $\vec{p}_f$
all quasiparticles travel with the same velocity, $\vf (\vec{p}_f)$.
In general there can be several branches of quasiparticles moving with the same
velocity but having different momenta.
Also the directions of $\vec{p}_f$ and $\vf (\vec{p}_f)$ are generally
different. However, for spherical or cylindrical Fermi surfaces 
$\vec{p}_f$ and $\vf (\vec{p}_f)$ differ only by a scaling factor.

The quasiclassical Green's function is solution of 
the following transport equation along a given trajectory,
and of the corresponding normalization 
condition\cite{eliashberg71,eilenberger68,larkin68,larkin75}
(the $\qt $-product is noncommutative and is explained in 
Appendix \ref{Notation})
\begin{equation}\label{qeq1}
[\check{\ep } - \check{h},
\check{g}]_\qt +i\hbar \qpartial \check{g}
= \check{0} ; \qquad
\check{g}\qt \check{g} = -\pi^2\, \check{1}.
\end{equation}
Here $\check{\ep } $ represents the energy variable, 
and $ \check{h}$ combines the molecular (or mean) field self-energies,
$\check{\sigma}_{mf}$, the impurity 
and electron-phonon self-energies, 
$\check{\sigma}_{i}$, 
and external potentials, $\check{v}_{ext}$
\beq
\check{h}= \check{\sigma}_{mf}+\check{\sigma}_{i}+\check{v}_{ext}
= \left( \begin{array}{cc} \hr & \hk \\ \tz & \ha
\end{array} \right), 
\eeq
which have the following explicit Keldysh structure
($\hat\tau_i$ denote Pauli matrices in the particle-hole space)
\ber
\check{\ep }=\left( \begin{array}{cc} \ep \tc & \tz \\ \tz & \ep \tc
\end{array} \right)&, \;&
\check{\sigma }_{mf}=\left( \begin{array}{cc} \smm_{mf} & \tz \\ \tz & \smm_{mf}
\end{array} \right), \; \\
\check{\sigma }_{i}=\left( \begin{array}{cc} \smmr & \smmk \\ \tz & \smma
\end{array} \right)&, \;&
\check{v }_{ext}=\left( \begin{array}{cc} \hat v_{ext} & \tz \\ \tz & \hat v_{ext}
\end{array} \right).
\eer
Disorder will be included by following standard averaging procedure
for dilute impurity concentrations.\cite{Abgo}
We denote impurity self-energies by $\check{\sigma}_{imp}$.
In quasiclassical approximation ($\mfp \gg 1/k_f$),
the impurity self-energy  
can be written in terms of the concentrations, $c_i$, of impurities
of type $i$ and the single impurity 
$t$-matrices, $\check t_i$,
\beq
\check{\sigma}_{imp} (\p_f ,\R,\ep ,t)=
\sum_{i=1}^{N} c_i \check{t}_i(\p_f,\p_f;\R, \ep, t).
\eeq
The $t$-matrices are solutions of the following equations
(we suppress the variables $\R, \ep, t$ for convenience)
\ber
\check{t}_i (\p_f,\p'_f)&=  &
\check{u}_i(\p_f,\p'_f) + 
\nonumber \\ 
&&N_f \Big\langle \check{u}_i(\p_f,\p''_f) 
\qt \check{g}(\p''_f) \qt \check{t}_i (\p''_f,\p'_f)
\Big\rangle_{\p''_f}\; , 
\eer
where $\check{u}_i(\p_f,\p''_f)$ is the scattering potential of an
impurity of type $i$.
The Fermi surface average $\langle \cdots \rangle_{\p''_f} $ is explained in
Appendix \ref{Notation}.
The impurity potential is diagonal in Keldysh space.

\section{Nambu-Gor'kov space structure}

We parameterize the elements of the Nambu matrices in the following way
\begin{eqnarray}
\gmra&=&\left( \begin{array}{cc} \plus \gaqra & \faqra \\ \fbqra & \gbqra
\end{array} \right) \; \; ,
\gmk=\left( \begin{array}{cc} \gaqk & \plus \faqk \\ -\fbqk & -\gbqk
\end{array} \right), \\
\hra &=&
 \left( \begin{array}{cc} \plus \vara & \Dara \\ \Dbra & \vbra
\end{array} \right) ,\;
\hk = 
 \left( \begin{array}{cc} \vak & \plus \Dak \\ -\Dbk & -\vbk
\end{array} \right)  .
\end{eqnarray}
Here $ \gaqra$, $ \fbqra$, $\Dak$ etc. are 2x2 spin matrices.

The molecular 
fields are determined by Landau's quasiparticle interaction function,
$A(\p_f,\p'_f)$, leading to a self-energy spin matrix,
$\hat \nu_{mf} (\p_F,\R,\ep ,t)$, which is diagonal in particle-hole
space. In superconductors this interaction must be supplemented by
the pairing interaction of quasiparticles, $V(\p_f,\p'_f)$,
which lead to an off-diagonal self-energy in particle-hole space,
$\hat \Delta_{mf}(\p_F,\R,\ep ,t)$. Thus,
\beq
\label{mf}
\smm_{mf}= \hat \nu_{mf} + \hat \Delta_{mf}.
\eeq
The mean-field self energies, Eq. (\ref{mf}), are
diagonal in Keldysh space.\cite{serene83} 
Their matrix structure in Nambu space is
\beq
\Dm_{mf}=\left( \begin{array}{cc} \plus 0 & \Da_{mf} \\ \Db_{mf} & 0
\end{array} \right), \;
\op\nu_{mf}=\left( \begin{array}{cc} \nu_{mf} & 0 \\ 0 & \tilde\nu_{mf}
\end{array} \right).
\eeq

Not all the matrix elements are independent from each other,
but are related by symmetry relations.\cite{serene83}
For instance, a quantity $x$ and the conjugated quantity $\tilde x$
are related by,
\beq
\label{tildesymm}
\tilde{x}(\p_f,\R,\ep,t)= x(-\p_f,\R,-\ep ,t)^{\ast}.
\eeq
The conjugation operator ($\tilde{\quad } $) defines an important transformation
of quasiclassical Green's functions and self-energies. We will use it
extensively in the following. 

\section{\State functions and Distribution functions}

The numerical solution of the transport equations for the quasiclassical Green's
functions can be simplified considerably by introducing a special
parameterization in terms of 2x2 spin matrix \state functions,
$\gara $, $\gbra$, and distribution functions,
$\xa $, and $\xb $, which transforms the original boundary
value problem for $\check g$ into initial value problems for the
\state and distribution spin matrices.
The normalization condition is in this formulation eliminated completely.
We present here the resulting equations
and refer for their derivation to
Appendices \ref{app_B} and \ref{app_C}. Before doing this we give a short
physical interpretation for the \state functions.
In the absence of particle-hole coherence, like in the equilibrium normal state,
 the functions $\gara$, $\gbra $ vanish. A superconductor, or a
normal metal in proximity to a superconductor, can be described in
equilibrium and in the clean limit by 
\name{Andreev}'s equations\cite{andreev64} 
with \name{Andreev} amplitudes $u$ and $v$.
Then, the \state function $\gar $, for example, is 
given in terms of the $u$- and $v$-spin matrices
(for positive energies) by the solution of the linear system
$\sum_{\beta } u_{\alpha \beta}\gar_{\beta \delta} = v_{\alpha \delta}$.
Thus, the \state functions are the transformation matrices between the
particle and hole like \name{Andreev} amplitudes. 
In the presence of quasiparticle damping
the \name{Andreev} description breaks down,
nevertheless one can {\it define } generalized amplitudes $u^\ra $ and $v^\ra $. In non-equilibrium they are defined by 
relations like $u^\ret \qt \gar = v^\ret $.  Note that these generalized
 amplitudes are defined by the quasiclassical Green's functions, 
not by wave functions. 

The quasiclassical Green's functions are conveniently parameterized by
\lrule
\ber 
\label{cgretav}
\gqra &=& \mp \, i\, \pi  \quad \Nra \qt \mat 
\left( 1 + \gara \qt \gbra \right) & 2\gara \\ -2\gbra &
-\left( 1 + \gbra \qt \gara \right) \matend \; ,\\[2mm]
\label{ckelgf2}
\gqk &=& -2\pi\, i  \quad \Nr \qt
\mat 
(\xa - \gar \qt \xb \qt \gba ) &-(\gar \qt \xb - \xa \qt \gaa ) \\
-(\gbr \qt \xa - \xb \qt \gba ) &(\xb - \gbr \qt \xa \qt \gaa ) \matend \qt
\Na ,\quad
\eer
with the `normalization matrices'
\ber
\label{cnormfac}
\Nra &=& \mat  (1-\gara \qt \gbra)^{-1} & 0 \\ 0 & (1-\gbra \qt \gara)^{-1} \matend \, \, .
\eer
In (\ref{cgretav}) the factor $\Nra$  may be written
on the left- or right-hand side.
The transport equations for the 2x2 spin matrix functions are
\ber
\label{cricc1}
\!\!\!\!
i\hbar \, \qpartial \gara +2\ep \gara &=& \gara \qt \Dbra \qt \gara + 
\Big( \vara \qt \gara - \gara \qt \vbra \Big) - \Dara \quad , \\[0.2cm]
\label{cricc2}
\!\!\!\!
i\hbar \, \qpartial \gbra -2\ep \gbra &=& \gbra \qt \Dara \qt \gbra + 
\Big( \vbra \qt \gbra - \gbra \qt \vara  \Big) - \Dbra \quad , \\[0.2cm]
\label{keld1}
\!\!\!\!
i\hbar \, \qpartial \xa + i\hbar \, \partial_t \xa &+&\Big(-\gar \qt \Dbr -\var\Big) \qt \xa +
\xa \qt \Big( -\Daa \qt \gba +\vaa \Big) = \nonumber \\[0.2cm]
&=& -\gar \qt \vbk \qt \gba
+\Dak \qt \gba + \gar \qt \Dbk -\vak \quad ,
\\[0.2cm]
\label{keld2}
\!\!\!\!
i\hbar \, \qpartial \xb - i\hbar \, \partial_t \xb &+&\Big( -\gbr \qt \Dar -\vbr\Big) \qt \xb +
\xb \qt \Big( -\Dba \qt \gaa + \vba \Big)= \nonumber \\[0.2cm]
&=& -\gbr \qt \vak \qt \gaa
+\Dbk \qt \gaa + \gbr \qt \Dak - \vbk \quad .
\eer
\rrule
\noindent
Equations (\ref{cgretav}), (\ref{cricc1}), and (\ref{cricc2})
generalize 
a useful formulation of the equilibrium theory
in terms of Riccati-type transport equations\cite{nagato93,schopohl95}
to non-equilibrium phenomena.
Equations (\ref{cgretav})-(\ref{keld2})
are new,\cite{eschrig97} numerically very stable and 
provide an efficient way to solve
non-equilibrium problems in superconductors.
Equations (\ref{cricc1})-(\ref{keld2})
need to be supplemented by initial conditions.
They are imposed for $\gar $, $\gba$, and $\xa $
at the beginning of the
trajectory, and for $\gaa $, $\gbr $, and $\xb $ 
at the end of the trajectory.
For correctly chosen initial conditions 
the transport equations for $\gar $, $\gba$, and $\xa $ are stable in
positive $\vec v_f $-direction, and the transport equations
for $\gaa $, $\gbr $, and $\xb $ are stable in
negative $\vec v_f $-direction.
In addition to the conjugation symmetries,
\ber
\gbra (\pf,\R,\ep,t)&=&\gara (-\pf,\R,-\ep,t)^{\ast} \;, \\
\xb(\pf,\R,\ep,t)&=& \xa(-\pf,\R,-\ep,t)^{\ast} \; ,
\eer
the \state and distribution functions obey the following symmetries
\ber
\gaa(\pf,\R,\ep,t)&=& \gbr(\pf,\R,\ep,t)^{\dagger} \; , \\
\xa (\pf,\R,\ep,t) &=& \xa (\pf,\R,\ep,t)^{\dagger}\; .
\eer
Note that the $\xa(\pf,\R,\ep,t) $, $\xb(\pf,\R,\ep,t) $ are hermitean spin
matrices.
In equilibrium,
\ber
\label{cxaF}
\xa_{eq} &=& \plus ( 1 - \gar \gba ) \tanh \frac{\ep}{2T} \; ,\\
\xb_{eq} &=& -( 1 - \gbr  \gaa ) \tanh \frac{\ep}{2T}.
\eer

\section{Explicit solution of Zaitsev's nonlinear
boundary conditions}

In the previous sections we have introduced a parametrization
of the non-equilibrium 
Keldysh Green's function, $\check g$, in terms of four \state functions and
two distribution functions (2x2 spin matrices)
\beq
\check g = \check g[\gar,\gbr,\gaa,\gba,\xa,\xb].
\eeq
An important problem is the formulation of boundary
conditions for these parameters  at surfaces and interfaces.\cite{zaitsev84,millis88,ashauer86,kieselmann87,nagai88,yip97}
A boundary condition for $\check g$
was obtained by Zaitsev,\cite{zaitsev84} which in principle
solves this problem for perfect interfaces. 
However, Zaitsev's non-linear boundary conditions have unphysical
spurious solutions which require special care, e.g. in a numerical
implementation.
A linearization of Zaitsev's boundary
conditions for the equilibirium was achieved recently by Yip\cite{yip97} for 
the case of an interface connected to infinite half spaces. 
Our solution generalizes these results to any interface geometry
and to non-equilibrium phenomena. 
Zaitsev's condition relates the
quasiclassical Green's functions with
Fermi velocity pointing in direction towards 
\begin{figure}[h]
\begin{minipage}{7.5cm}
\centerline{
\epsfxsize8.0cm
\epsffile{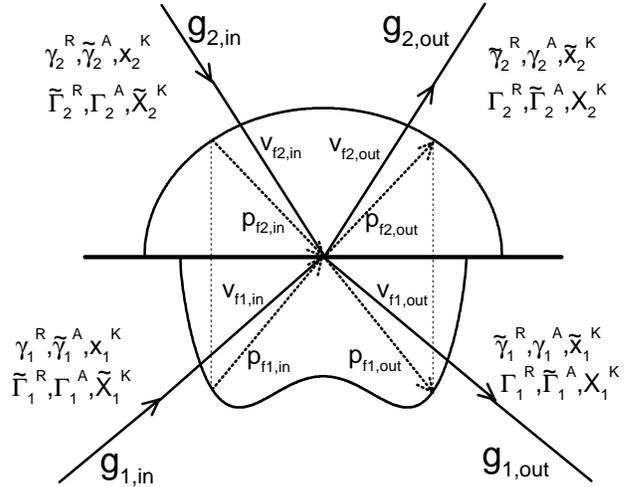}
}
\end{minipage}\\[0.5cm]
\begin{minipage}{8.5cm}
\caption{ \label{zaitsev}
Notation for the Green's functions at the
interface. Indices 1 and 2 refer to the sides of the interface.
The arrows for the Fermi momenta (dotted) are for particle like excitations.
The Fermi velocity directions (full lines) are given by the directions
perpendicular to the Fermi surface (full curves)
at the corresponding Fermi momentum.
The components of the Fermi momenta parallel to the surface are conserved
(indicated by the thin dotted lines).
}
\end{minipage}
\end{figure}
\noindent 
the surface, $\check g_{1,in}$,
$\check g_{2,in}$, and those with Fermi velocity
pointing away, $\check g_{1,out}$, $\check g_{2,out}$. 
Indices 1 and 2 refer to the two sides of the interface (see
Fig. \ref{zaitsev}). 
Using the definitions
\ber
\check P_1&=&\frac{i}{2\pi } \left( \check g_{1,in} + \check g_{1,out} \right) ,\quad
\check P_2=\frac{i}{2\pi } \left( \check g_{2,in} + \check g_{2,out} \right), \\
\check P_a&=&\frac{i}{2\pi } \left( \check g_{1,in}-\check g_{1,out}\right) =
\frac{i}{2\pi } \left( \check g_{2,out}-\check g_{2,in}\right) ,
\eer
which fulfill the relations
\ber
\check P_a \qt \check P_1+\check P_1 \qt \check P_a&=&\check 0\; , \quad
\check P_a \qt \check P_a + \check P_1 \qt \check P_1 = \check 1 \; , \\
\check P_a \qt \check P_2+\check P_2 \qt \check P_a&=&\check 0\; , \quad
\check P_a \qt \check P_a + \check P_2 \qt \check P_2 = \check 1\; ,
\eer
Zaitsev's boundary conditions read\cite{zaitsev84}
\ber
\label{zaitsevbc}
&&\left[ \left( \check 1 - \check P_a \right) \qt \check P_1 \qt \check P_2
- \left( \check 1 + \check P_a \right) \qt \check P_2 \qt \check P_1 \right]
(1-{\cal R}) = \nn \\
&&=-2\check P_a \qt (\check 1 - \check P_a \qt \check P_a ) (1+{\cal R}).
\eer
Here, and in the following ${\cal R}={\cal R}(\vec{p}_f )$ and 
${\cal D}={\cal D}(\vec{p}_f )$ are the reflection
and transmission coefficients of the interface for quasiparticles
in the normal state correspondingly, 
${\cal R}(\vec{p}_f )+{\cal D}(\vec{p}_f )=1$.
In the following we present the explicit solution of
Zaitsev's nonlinear boundary conditions at
spin-conserving interfaces in terms of 
the \state functions and distribution functions introduced above.
It is useful for this purpose to introduce a notation which indicates
the stability properties of solutions of the transport equations.
We use capital letters ($\Gara,
\Gbra, \Xa, \Xb $) for functions, which are
stable solutions when integrating the transport equation towards the
surface. Small case letters ($\gara, \gbra ,\xa,\xb $) 
are used for functions, which are stable in the direction away from
the surface. We also generalize the notation for the conjugation
operation. It includes an conversion from small case to capital case
letters,
\beq
\label{tildesymm1}
\tilde{x}(\p_f,\R,\ep,t)= X(-\p_f,\R,-\ep ,t)^{\ast}.
\eeq
By integrating in direction towards the surface, the quantities 
$\gara_j,\gbra_j,\xa_j,\xb_j$, ($j=1,2$) are known.
The quantities
$\Gara_j,\Gbra_j,\Xa_j,\Xb_j$
are to be determined by integrating in direction away from the surface.
At the surface the second set of quantities is determined in terms of
the first one by boundary conditions.

The incoming quasiclassical retarded Green's functions (with velocity direction
towards the interface) on each side of the
interface are given then by (see Fig. \ref{zaitsev}),
\ber
\label{g1in}
\check g_{1,in}&=&\check g [\gar_1,\Gbr_1,\Gaa_1,\gba_1,\xa_1,\Xb_1] \; ,\\
\label{g2in}
\check g_{2,in}&=&\check g [\gar_2,\Gbr_2,\Gaa_2,\gba_2,\xa_2,\Xb_2] \; ,
\eer
and the outgoing ones (with velocity direction away from the interface)
\ber
\label{g1out}
\check g_{1,out}&=&\check g [\Gar_1,\gbr_1,\gaa_1,\Gba_1,\Xa_1,\xb_1] \; ,\\
\label{g2out}
\check g_{2,out}&=&\check g [\Gar_2,\gbr_2,\gaa_2,\Gba_2,\Xa_2,\xb_2] \; .
\eer
Using our parametrization, Zaitsev's boundary conditions
can be solved for the unknown quantities in a straightforward
way. 
In the superconducting state
we define effective reflection and transmission coefficients, which
we present in Appendix \ref{refcoeff}.
The sum of each generalized reflection coefficient with its corresponding
transmission coefficient is equal to one.
Using these coefficients we can write the general boundary conditions
for the six unknown spin matrix distributions functions in a compact form.
For the \state functions we have\cite{footn}
\lrule
\ber
\label{bcc1}
\Gara_1 &=& R_{1l}^{\ra} \qt \gara_1 + D_{1l}^{\ra} \qt \gara_2 =
\gara_1 \qt R_{1r}^{\ra} + \gara_2 \qt D_{1r}^{\ra} \quad ,
\\
\label{bcc2}
\Gbra_1 &=& \tilde R_{1l}^{\ra} \qt \gbra_1 + \tilde D_{1l}^{\ra} \qt \gbra_2 = 
\gbra_1 \qt \tilde R_{1r}^{\ra} + \gbra_2 \qt \tilde D_{1r}^{\ra} \quad .
\eer
Note the intuitively appealing structure of the relations. The outgoing 
functions are
weighted averages of two incoming functions. The
weights depend on the incoming parameters as well, which reflects
the coherence during Andreev reflection. The distribution functions
have the following boundary conditions
\ber
\label{bck1}
\Xa_1 &=& 
\frac{R_{1l}^{\ret}}{\cal R} \qt {\cal R} \xa_1 \qt \frac{\tilde R_{1r}^{\adv}}{\cal R}
+ \frac{D_{1l}^{\ret}}{\cal D} \qt {\cal D} \xa_2 \qt \frac{\tilde D_{1r}^{\adv}}{\cal D}
- A_{1l}^{\ret}\qt {\cal RD}\xb_2 \qt \tilde A_{1r}^{\adv} 
\quad ,
\\
\label{bck2}
\Xb_1 &=& 
\frac{\tilde R_{1l}^{\ret}}{\cal R}\qt {\cal R} \xb_1 \qt \frac{
R_{1r}^{\adv}}{\cal R}
+ \frac{\tilde D_{1l}^{\ret}}{\cal D}\qt {\cal D} \xb_2 \qt \frac{
D_{1r}^{\adv}}{\cal D}
- \tilde A_{1l}^{\ret} \qt {\cal RD}\xa_2 \qt A_{1r}^{\adv} \quad .
\eer
\rrule
Analogous relations, obtained by interchanging the subscripts 1 and 2,
hold for the other side of the interface. 
The terms proportional to the product ${\cal RD}={\cal D}
(1-{\cal D})$, are due to
particle-hole interference and
do not arise in the classical limit. 
Insertion of 
these equations into Zaitsev's boundary conditions
shows, that they solve the nonlinear problem and eliminate
all spurious solutions.
\section{Linear Response theory}

The general linear response theory in terms of the \state functions and
distribution functions was
developed in Refs. [\onlinecite{eschrig97,eschrig98}].
Here we give a short review of the relevant equations and generalize them
for spin dependent phenomena.
For the special case of the diamagnetic response
see Belzig, Bruder, and Fauch{\`e}re.\cite{belzig98}
We assume a small external perturbation and expand
$\check{g}$ and $\check{h}$ around the unperturbed solutions.
With the replacements $\check g \to \check g + \delta \check g$ and
$\check h \to \check h + \delta \check h$ we arrive in linear order
at the equations
\ber \label{lrqeq1}
[\check{\ep } - \check{h},
\delta \check{g}]_\qt +i\hbar \qpartial \delta \check{g}
&=& [\delta \check{h},\check{g}]_\qt \quad , \\
\label{lrnorm}
\delta \check{g}\qt \check{g} + \check{g} \qt \delta \check{g} &=& \check{0} \; .
\eer
Here the linearized self consistency equations determine $\delta \check h$.
For a specially chosen parametrization 
given at the end of Appendices \ref{app_B} and \ref{app_C},
the linear correction of the Green's function, $\delta \check g$,
can be written in terms of the linear corrections to the \state
functions, $\dgara$, $\dgbra $, and linear corrections to
the distribution functions, $\dxak $, $\dxbk $. 

It is convenient to transform from the Keldysh response, 
$\dgqk $ to the {\em anomalous response }, $\dgqan $,
\beq
\dgqan = \dgqk - \dgqr\qt \Fe + \Fe \qt \dgqa \; ,
\eeq
with $F_{eq}= \tanh \epsilon/2T $.
Then, with the definition of the anomalous components of the
distribution spin matrices,
\ber
\label{xano}
\dxa &=& \dxak + \gar \qt \Fe \qt \dgba + \dgar \qt \Fe \qt \gba \; ,\\
\dxb &=& \dxbk - \gbr \qt \Fe \qt \dgaa - \dgbr \qt \Fe \qt \gaa \; ,
\eer
the {\it spectral response}, $\dgqra $, and the anomalous
response, $\dgqan $, are given by
\lrule
\begin{equation}
\label{linretadv}
\dgqra
=\mp 2\pi i \; \Nra \; \qt \; \left(
\begin{array}{cc}
(\dgara \qt \gbra + \gara \qt
\dgbra ) & (\dgara  + \gara\qt
\dgbra \qt \gara) \\
-(\dgbra + \gbra \qt
\dgara  \qt \gbra ) &
-(\dgbra \qt \gara +  \gbra\qt
\dgara ) \end{array} \right) \; \qt \; \Nra \quad ,
\end{equation}
\begin{equation}
\label{linano}
\dgqan
=- 2\pi i \; \Nr \; \qt \; \left(
\begin{array}{cc}
(\dxa- \gar \qt
\dxb  \qt  \gba) & 
-(\gar \qt \dxb  - \dxa \qt \gaa) \\
-(\gbr\qt \dxa - \dxb  \qt
\gba ) &
(\dxb -  \gbr\qt
\dxa  \qt  \gaa) \end{array} \right) \; \qt \; \Na
\; .
\end{equation}
\rrule
Using the anomalous self-energies,
\beq
\dhan = \dhk - \dhr\qt \Fe + \Fe \qt \dha \; ,
\eeq
we define the following short-hand notation for the driving terms in
the transport equations,
\begin{eqnarray}\label{deforder1}
\dhra =
{\left(
\begin{array}{cc}
\dvara &\dDara\\
\dDbra& \dvbra
\end{array}
\right)}, \ \dhan ={\left(
\begin{array}{cc}
\plus \dvaan&\plus \dDaan\\
-\dDban& -\dvban
\end{array}\right)}\ .
\end{eqnarray}
The spin matrices $\dgara $, $\dgbra $, $\dxa $, and $\dxb $ 
are functions of $\vec{p}_f$, $\vec{R}$, $\epsilon$,
$t$, and satisfy the transport equations
\lrule
\begin{eqnarray}\label{transpqcls1}
i\hbar \, \qpartial \dgara+2\epsilon
\dgara
&-&(\gara \Dbra +\vara)
\qt\dgara
+\dgara\qt(-\Dbra\gara +
\vbra ) \nonumber\\
&=& \gara \qt\dDbra \qt\gara
\!\!+\dvara\qt\gara\!\!-\gara\qt
\dvbra -\dDara\! , 
\end{eqnarray}
\begin{eqnarray}
\label{transpqcls2}
i\hbar \, \qpartial \dgbra-2\epsilon
\dgbra
&-&(\gbra\Dara+\vbra)
\qt\dgbra
+\dgbra \qt(-\Dara \gbra +
\vara ) \nonumber\\
&=&
\gbra \qt\dDara\qt\gbra
\!\!+\dvbra\qt\gbra -
\gbra \qt
\dvara\!\! -\dDbra \! ,
\end{eqnarray}
\begin{eqnarray}
i\hbar \, \qpartial \dxaan  +i\hbar \, \partial_t\dxaan
&+& (-\gar \Dbr \!\!-\var )\qt
\dxaan +\dxaan \qt (-\Daa \gba
+\vaa )\nonumber\\ \label{transpqcls3}
& =&
-\gar \qt\dvban \qt \gba
+\dDaan \qt\gba +\gar \qt
\dDban -\dvaan \ ,
\end{eqnarray}
\begin{eqnarray}
i\hbar \, \qpartial \dxban
-i\hbar \partial_t \dxban
&+& (-\gbr \Dar \!\!-\vbr )\qt
\dxban+\dxban \qt (-\Dba \gaa
+\vba)\nonumber\\ \label{transpqcls4}
& =&
-\gbr \qt \dvaan \qt\gaa
+\dDban \qt\gaa +\gbr \qt
\dDaan -\dvban
\ .
\end{eqnarray}
\rrule
One convenient feature of our parameterization is the fact,
that the linear response transport equations 
(\ref{transpqcls1})-(\ref{transpqcls4}) decouple for given self-energies.
Furthermore, the transport equations for 
$\dgar $, $\dgba $, $\dxaan $ are stable
in direction of $\vec{v}_f$, and the transport equations for
$\dgbr $, $\dgaa $, 
$\dxban $ are stable in direction of $-\vec{v}_f$. 
This makes a numerical
treatment much easier than solving the boundary value problem
for the coupled transport equations (\ref{lrqeq1})-(\ref{lrnorm}).
The $\vec{R}$-points for the initial condition correspond to
the final point or the initial point of the trajectory depending on the
direction of stability of the transport equation.

Finally we present the boundary conditions for the
\state functions and for the distribution functions in linear response.
With an analogous definition of the anomalous components of the
outgoing distribution spin matrices,
\ber
\dXa &=& \dXak + \Gar \qt \Fe \qt \dGba + \dGar \qt \Fe \qt \Gba \; ,\\
\dXb &=& \dXbk - \Gbr \qt \Fe \qt \dGaa - \dGbr \qt \Fe \qt \Gaa \; ,
\eer

we obtain the boundary conditions for the corrections to the
\state functions and distribution functions,
\lrule
\ber
\dGar_1 &=& 
\frac{R_{1l}^{\ret}}{\cal R} \qt {\cal R} \dgar_1 \qt \frac{R_{1r}^{\ret}}{\cal R}
+ \frac{D_{1l}^{\ret}}{\cal D} \qt {\cal D} \dgar_2 \qt \frac{D_{1r}^{\ret}}{\cal D}
+ A_{1l}^{\ret}\qt {\cal RD}\dgbr_2 \qt A_{1r}^{\ret}  \quad ,
\\
\dGbr_1 &=& 
\frac{\tilde R_{1l}^{\ret}}{\cal R} \qt {\cal R} \dgbr_1 \qt \frac{\tilde R_{1r}^{\ret}}{\cal R}
+ \frac{\tilde D_{1l}^{\ret}}{\cal D} \qt {\cal D} \dgbr_2 \qt \frac{\tilde D_{1r}^{\ret}}{\cal D}
+ \tilde A_{1l}^{\ret}\qt {\cal RD}\dgar_2 \qt \tilde A_{1r}^{\ret} \quad , 
\eer

\ber
\dXa_1 &=& 
\frac{R_{1l}^{\ret}}{\cal R} \qt {\cal R} \dxa_1 \qt \frac{\tilde R_{1r}^{\adv}}{\cal R}
+ \frac{D_{1l}^{\ret}}{\cal D} \qt {\cal D} \dxa_2 \qt \frac{\tilde D_{1r}^{\adv}}{\cal D}
- A_{1l}^{\ret}\qt {\cal RD}\dxb_2 \qt \tilde A_{1r}^{\adv} \quad ,
\\
\dXb_1 &=& 
\frac{\tilde R_{1l}^{\ret}}{\cal R} \qt {\cal R} \dxb_1 \qt \frac{ R_{1r}^{\adv}}{\cal R}
+ \frac{\tilde D_{1l}^{\ret}}{\cal D} \qt {\cal D} \dxb_2 \qt \frac{ D_{1r}^{\adv}}{\cal D}
- \tilde A_{1l}^{\ret}\qt {\cal RD}\dxa_2 \qt A_{1r}^{\adv} \quad .
\eer
\rrule

\section{Andreev spectroscopy at N-S interfaces for unconventional
superconductors}

To illustrate the physical content of the introduced distribution functions 
we discuss in this section the Andreev reflection process at 
an interface between a normal metal (subscript `1') and a $d$-wave
superconductor (subscript `2'). 
This problem was studied by Blonder, Tinkham, and Klapwijk\cite{blonder82} 
for conventional $s$-wave superconductors, and was generalized
to unconventional superconductors by Bruder.\cite{bruder90}
We generalize these calculations to include finite impurity scattering
and identify features which are
sensitive to time reversal symmetry breaking states.
In an Andreev reflection  experiment
a beam of
normal quasiparticles with energies,
$\epsilon_b$,  and momenta, $\vec{p}_{f,b}$,
is injected across the interface into the superconductor. 
 Two types of reflections will occur.
Part of the beam will be  regularly  reflected  at the interface,
which amounts to a reflection of the quasiparticle's velocity, momentum 
and  current, and part will be Andreev reflected. 
Andreev's retro-reflection 
is caused by particle-hole conversion which reverses the velocity but
conserves momentum and
current to very good approximation. Because the current is affected
quite differently by regular reflection and Andreev reflection, a
measurement of the current-voltage characteristics provides 
direct information on the balance between these two reflection processes.
Together with a  thorough theoretical analysis such measurements
inform about fundamental properties of the superconductor such as
the symmetry of pairing,\cite{vanharlingen95} the gap size and anisotropies, 
and interface resonance states.\cite{buchholtz81,hu94,matsumoto95,tanaka96,fogelstrom97,alff97}
For anisotropic 
superconductors both the current density in the
reflected and the Andreev reflected beams
will depend  strongly on the direction of the incoming beam, in
addition to their dependence of the energy of the incoming quasiparticles.

The following calculation of Andreev reflection includes anisotropic
pairing, a finite mean free path in the superconductor, 
a finite transparency of the N-S interface, the layer of
a strongly distorted order parameter near the interface, and the
effects of the interface on the excitation spectrum, in particular
the low-energy bound states.
We consider a layered $d$-wave superconductor with cylindrical
Fermi surface and isotropic Fermi velocity along the layers. 
The interface lies 
perpendicular to the layers and we  assume, for simplicity, the same
Fermi velocity in the normal  and the superconducting parts of
the N-S contact.

The \state functions
$\gar_1$, $\gba_1$ are determined by boundary conditions at infinity,
Eqs. (\ref{hom1}), (\ref{hom2}),
whereas $\Gbr_1$, $\Gaa_1$ are determined by the interface boundary conditions,
Eqs. (\ref{bcc1}), (\ref{bcc2}).
For the spin singlet superconductor we write $\gar = i\sigma_y \ga $,
$\gbr = i\sigma_y \gb $, $\Gar = i\sigma_y \Ga $, and
$\Gbr = i\sigma_y \Gb $, where $\ga$, $\gb$, $\Ga$, $\Gb $ are scalar functions.
On the normal side the incoming \state functions
$\ga_1 $, $\gb_1 $ are zero as a consequence of their zero initial values at
infinity.
Thus, the retarded part of the Green's function on the normal side, following
from Eqs. (\ref{g1in}), (\ref{g1out}), and (\ref{cgretav}), has the form
\ber
\label{struc_e}
\gqr_{1,in}&=& 
- i \pi \; \left( \begin{array}{cc}
1 &  \plus 0 \\ -2i\sigma_y \Gb_1 & - 1 \end{array} \right)
\; ,\\
\label{struc1_e}
\gqr_{1,out}&=& 
- i \pi \; \left( \begin{array}{cc}
1 &  2i\sigma_y \Ga_1 \\ 0 & - 1 \end{array} \right)
\; .
\eer
The nonzero quantities $\Ga_1 $ and $\Gb_1$, 
describe the proximity effect at the N-S interface. 
The solutions for $\Ga_1$ and $\Gb_1$ in the normal metal in
equilibrium are
\ber
\Ga_1(x,\epsilon ) &=& \Ga_{1}(\epsilon) e^{i\frac{2\epsilon }{v_f}x}
e^{-\frac{x}{v_f \tau_1 }} ,\\
\Gb_1(x,\epsilon ) &=& \Gb_{1}(\epsilon) e^{-i\frac{2\epsilon }{v_f}x}
e^{\frac{x}{v_f \tau_1 }},
\eer
where the spatial trajectory 
coordinate $x$ is measured in direction of $\vf $ and is zero at the interface,
positive for $\Ga_1$ and negative for $\Gb_1$, 
and $\tau_1 $ is the lifetime in the normal metal. Both amplitudes decay
from the interface towards the normal metal on a scale $\vf \tau_1$.
For simplicity, 
we assume in all what follows that the normal metal is in the clean limit.
The Tomasch oscillation factors,\cite{Tomasch66} 
with Tomasch wave length $\pi v_f/\epsilon $,
are carried by $\Ga_{1}$, $\Gb_{1}$, whereas $\gar_2$ and $\gba_2$
vary only in the region of varying order parameter near the interface and
are constant far away. Similarly, on the superconducting side, far away from
the interface, the deviations of the outgoing \state functions from 
their homogeneous solutions, 
$\Ga_{2}(x )-\Ga_{2,hom}$, $\Gb_{2}(x )-\Gb_{2,hom}$,
carry the Tomasch oscillations with
wavelength $\pi v_f/ \sqrt{\epsilon^2-|\Delta |^2}$ if $|\epsilon | > 
|\Delta | $.
In the following all quantities without 
spatial argument refer to their values at the interface.

In quasiclassical approximation the 
incoming beam of non-equilibrium excitations with energy, $\epsilon_b$,
and momentum, $\vec{p}_{f,b}$, is described by  the 
``scattering'' part of the Keldysh Green's function, $\Delta \gqk = 
\gqk -\gqk_{eq}$, where
the equilibrium Keldysh Green's function $\gqk_{eq}$ is subtracted.
In the following we assume, for simplicity, 
a spin unpolarized incoming beam.
The calculations for spin-polarized beams pose no new problems
but are of interest only for high-field
superconductivity,\cite{meservey94}
spin-triplet 
pairing,\cite{kieselmann83,yip85,zhang88,kurkijarvi90} contacts between
superconductors and magnetic materials,\cite{tokuyasu88}
or spin-active interfaces.\cite{millis88,tokuyasu88}
The incoming beam is then characterized by unit spin-matrix
distribution functions $\Delta \xa_1$ and $\Delta \Xb_1$.
To obtain a physical interpretation of this distribution functions we
consider a solution of Eq. (\ref{keld1}) in form of a traveling wave with frequency $\omega $,
\beq
\Delta \xa_1 (x,\epsilon,t) =
\Delta \xa_{1} (\epsilon) e^{i\frac{\omega }{v_f}(x-v_f t)}.
\eeq
The corresponding part of the Keldysh Green's function 
follows from
Eq. (\ref{ckelgf2}), and after performing the time convolutions, Eq.
(\ref{timeconv}), we obtain
\lrule
\ber
&&\Delta \gqk_{1,in} (x,\epsilon, t)
= - 2\pi i \\
&&\times \Bigg\{ \Delta \xa_{1}(\epsilon )
\left( \begin{array}{cc}
e^{i\frac{\omega }{v_f}(x- v_f t)}
&  -i\sigma_y \Gb_{1}^\ast(\epsilon-\frac{\omega }{2})
e^{i\frac{2\epsilon}{v_f}(x-\frac{\omega }{2\epsilon }v_f t)}
\\
-i\sigma_y \Gb_{1} (\epsilon+\frac{\omega }{2})
e^{-i\frac{2\epsilon}{v_f}(x+\frac{\omega }{2\epsilon }v_f t)}
&
- \Gb_{1}(\epsilon+\frac{\omega }{2})
\Gb_{1}^\ast (\epsilon-\frac{\omega }{2})
e^{-i\frac{\omega }{v_f}(x+ v_ft)}
\end{array} \right)
+\left( \begin{array}{cc}
0 & 0 \\ 0 & \Delta \Xb_{1}(x,\epsilon ,t) \end{array} \right)
\Bigg\}
\enspace . \nonumber 
\eer
\rrule
This gives us a very transparent interpretation for the processes covered
by $\Delta \xa_1$.
The upper left entry describes an incoming particle with velocity
$v_f$. The lower right entry describes an Andreev reflected hole
with velocity $-v_f$, coming from the interface due to
retro-reflection combined with particle hole conversion. 
The off-diagonal components describe particle- and hole-like
Tomasch oscillations due to particle-hole coherence. 
The degree of coherence between particles and holes in the incoming 
distribution, $\Delta \xa_1$, is given by the \state function $\Gb_1$.
This gives a direct physical interpretation of the \state functions.
Similarly, $\Ga_1 $ is the amplitude
for Andreev reflected particles due to an incoming hole excitation beam.
On the other hand, the distribution function $\Delta \Xb_1$
describes an incoherent hole coming from the interface. This
component can be nonzero only if there is an incoming hole in the
Green's function $\Delta \gqk_{1,out} $ or
$\Delta \gqk_{2,out} $, which we exclude in our
scattering boundary condition. 
Thus, the correct boundary conditions for the scattering problem take
the intuitively appealing form,
to allow for the incoming particle beam only an incoming distribution
function, $\Delta \xa_1 $, and for all outgoing channels only
outgoing distribution functions, $\Delta \Xa_1 $, $\Delta \Xa_2 $,
$\Delta \Xb_2$.
All other distribution function components are zero.

In the following we assume a stationary ($\omega = 0$)
situation, where an incoherent beam is injected, which allows us to consider
the incoming beam spatially homogeneous along the trajectory.
Furthermore, it is sufficient to solve the problem for the distribution function
\beq
\Delta \xa_1 = 
 -8 \pi \delta \epsilon \; 
\delta(\epsilon-\epsilon_b ) \delta(\hat\vec{p}_f-\hat\vec{p}_{f,b}
)\enspace,
\eeq
where $\hat\vec{p}_f$ denotes a unit vector in direction $\vec{p}_f$, and
$\delta \epsilon $ is the energy resolution of the beam. Any other
distribution of incoming excitations is then given by a linear combination
of such solutions with properly chosen weight functions.
The  current density of the incoming beam is $ j_0=2e N_f v_f \delta \epsilon $.
For a  current density much smaller  than the critical current
density 
in the superconductor, one can neglect the  
effect of the beam on the self-consistent  order 
parameter and impurity self-energies.

For the scattering parts of the Keldysh Green's function at the normal side
we obtain
\ber
\label{GAnd}
\Delta \gqk_{1,in}
&=&- 2\pi i \; \Delta \xa_1 
\; \left( \begin{array}{cc}
1 &  -i\sigma_y \Gb_1^\ast \\ -i\sigma_y \Gb_1 & -  | \Gb_1 |^2 \end{array} \right)\enspace
, \\
\label{Gref}
\label{ano_Refl}
\Delta \gqk_{1,out} &=&- 2\pi i \; \Delta \Xa_1 
\; \left( \begin{array}{cc}
1 &  0 \\ 0 & 0 \end{array} \right)\enspace . 
\eer
The vanishing off-diagonal elements
of the reflected Green's function show
that there is no hole admixing in the reflected particle beam. 

The boundary conditions for the N-S interface follow from
Eqs. (\ref{bcc1})-(\ref{bck2}), 
\ber
\label{bccxa1}
&&\Delta \Xa_1 = 
{\cal R} \left|\frac{1+\ga_2 \gb_2 }{1+{\cal R} \ga_2 \gb_2 }\right|^2 \Delta \xa_1 ,\quad \\
\label{bccxa2}
\Delta \Xa_2 &=& {\cal D} \Delta \xa_1 ,\quad \quad \; \;
\Delta \Xb_2 = -{\cal RD} \left| \gb_2 \right|^2 \Delta \xa_1 \quad ,
\\
\label{bccga1}
\Ga_1 &=& {\cal D} \frac{ \ga_2 }{1+{\cal R} \ga_2 \gb_2 }, \qquad
\Ga_2 = {\cal R} \ga_2 , 
\\
\label{bccgb1}
\Gb_1 &=& {\cal D} \frac{ \gb_2 }{1+{\cal R} \ga_2 \gb_2 },
\qquad \Gb_2 = {\cal R}  \gb_2 \quad .
\eer

The total current densities are given in terms of the Keldysh Green's functions
via the formula
$\vec{j}= eN_f \int (d\epsilon /8\pi i ) \mbox{Tr}
\langle \tau_3 \vec{v}_f \Delta \gqk \rangle $.\cite{serene83}
Using the boundary conditions (\ref{bccxa1})-(\ref{bccgb1}),
this gives directly the total current densities at the interface
in terms of the injected current density,
\ber
\label{rand}
\frac{j_{1,in}}{j_0} &=&  1+ {\cal D}^2 
\left| \frac{\gb_2}{1+{\cal R} \gb_2 \ga_2 } \right|^2  \; ,\\
\label{rcon}
\frac{j_{1,out}}{j_0} &=& {\cal R} \left| \frac{1+\ga_2 \gb_2 }{1+{\cal R} \ga_2 \gb_2 } \right|^2 \; , \\
\label{tand}
\frac{j_{2,in}}{j_0} &=&  {\cal RD} 
\frac{\left| \gb_2 \right|^2 (1+\left| \ga_2 \right|^2) }{\left| 1+ {\cal R}\gb_2 \ga_2 \right|^2 }  \; , \\
\label{tcon}
\frac{j_{2,out}}{j_0} &=& {\cal D} \frac{1+\left| \gb_2 \right|^2}{\left| 1+{\cal R} \ga_2 \gb_2 \right|^2 } \; .
\eer
Here, $j_{1,in} $ describes the incoming current including the excess current,
$j_{2,in} $ the regularly reflected current, $j_{2,out}$ the 
regularly transmitted current, and
$j_{2,in}$ describes the process where the Andreev reflected holes
are regularly reflected back to the superconductor at the interface.
For energies below the gap the transmitted current densities, $j_{2,in}$,
$j_{2,out}$,
decay with distance from
the interface into the superconducting region, where they are converted into
super-currents.
It is straightforward to show
the conservation law $j_{1,in}+j_{2,in}=j_{1,out}+j_{2,out}$.
Eqs. (\ref{rand})-(\ref{tcon}) hold for general anisotropic and
unconventional superconductors, including impurity scattering.
The quantities $\ga_2$ and $\gb_2$ 
follow from solving numerically
their transport equations, Eqs. (\ref{cricc1}) and (\ref{cricc2}), 
with self consistently 
determined self-energies and order parameter.  For 
\begin{figure}[h]
\begin{minipage}{8.2cm}
\begin{minipage}{8.1cm}
\centerline{
\epsfxsize8.5cm
\epsffile{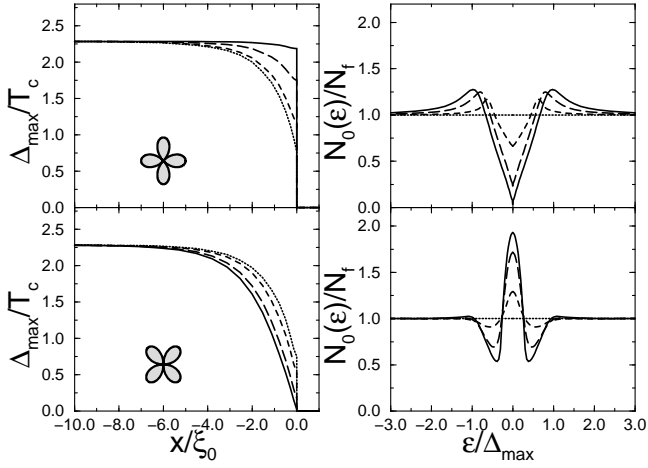}
}
\end{minipage}\\
\begin{minipage}{8.1cm}
\caption{ \label{gaps_dos_0.1}
Order parameter amplitude (left) and local density of states at
the interface (right) for a interface in (100) direction (top) and
in (110) direction (bottom). The interface is at $x=0$, the normal metal
extends to $x>0$.
The transmission coefficients for the
different curves are ${\cal D}_0=0.1 $ (full line), ${\cal D}_0=0.5 $
(long dashed), ${\cal D}_0=0.9 $ (dashed), and ${\cal D}_0=1.0 $ (dotted).
The temperature is $T=0.3T_c$, and the mean free path $\ell =10 \xi_0$.
}
\end{minipage}
\end{minipage}
\end{figure}
\noindent
conventional $s$-wave superconductors, and assuming 
a step function for the order parameter our formulae agree with the
results of 
Blonder, Tinkham and Klapwijk.\cite{blonder82}
It is clear from Eq. (\ref{rand}) that the Andreev reflected beam 
always enhances 
the current density in the injection beam, giving
rise to the excess current.
The enhancement is proportional to ${\cal D}^2$,
reflecting the fact that both the incoming particle and the Andreev reflected
hole have to cross the interface.
On the other hand,
the current density of the conventionally
reflected beam, described by Eq. (\ref{rcon}),
can be below or above the value ${\cal R} \cdot j_0$.

The angle resolved density of states at the superconducting side of
the interface is given by
\beq
\label{DOS}
N (\epsilon, \vec{p}_f ) = 
N_f \Re \frac{1-{\cal R}\ga_2 \gb_2 }{1+{\cal R } \ga_2 \gb_2  } .
\eeq
The local density of states is given by the Fermi surface average over this
expression.
Eq. (\ref{DOS}) shows that interface bound states are given by the solution
of the equation $1+{\cal R } \ga_2 \gb_2 =0$. Because the absolute values of
$\ga_2$ and $\gb_2$ are in equilibrium always smaller than or equal to 
unity, bound states at an interface can strictly occur only for
${\cal R}=1$, that means zero transmission. For finite transmission the
bound states broaden into interface resonances.
Impurity scattering further broadens these resonances. 
In Fig. \ref{gaps_dos_0.1} we show our self consistent solutions for
the $d$-wave order parameter, $\Delta = \sqrt{2} \cos 2\psi $,
and for the local density of quasiparticle states at the interface.
For definiteness, we modeled the angular dependence of the
transmission coefficient for the N-N interface by,
\beq
{\cal D}(\phi )=\frac{{\cal D}_0\sin^2 \phi }{{\cal R}_0 + {\cal D}_0 \sin^2 \phi } 
\; ,
\eeq
appropriate for a $\delta $-function potential barrier.
Here, $\phi $ is the impact angle between incoming trajectory and
interface. The parameters ${\cal D}_0$ and
${\cal R}_0=1-{\cal D}_0$ are the transmission and reflection
coefficients for perpendicular impact ($\phi=\pi/2$).
The impurity self-energy was calculated self consistently in Born
approximation with a 
\lrule
\begin{figure}[h]
\begin{minipage}{16.3cm}
\begin{minipage}{8.1cm}
\centerline{
\epsfxsize8.0cm
\epsffile{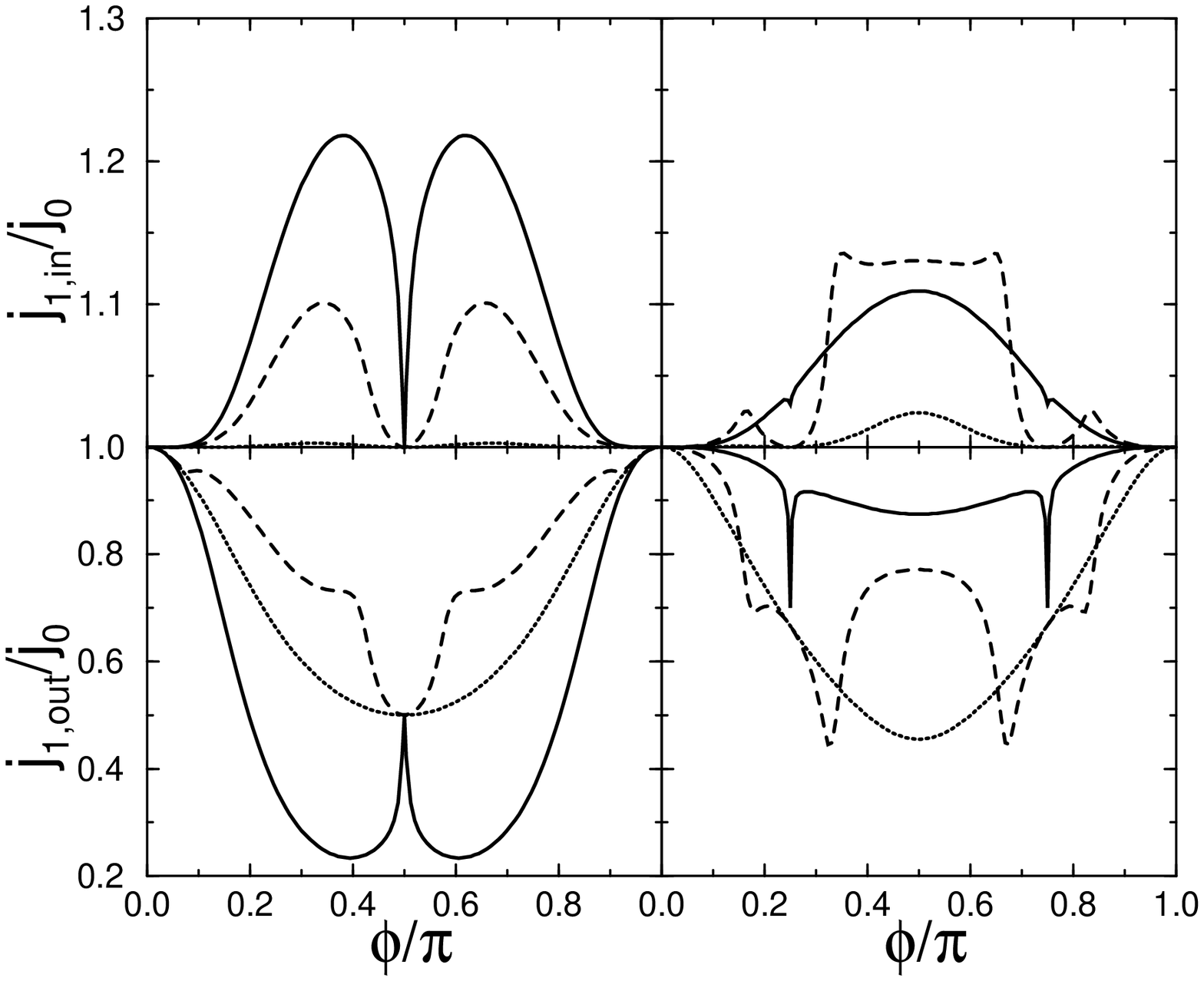}
}
\end{minipage}
\begin{minipage}{8.1cm}
\centerline{
\epsfxsize8.0cm
\epsffile{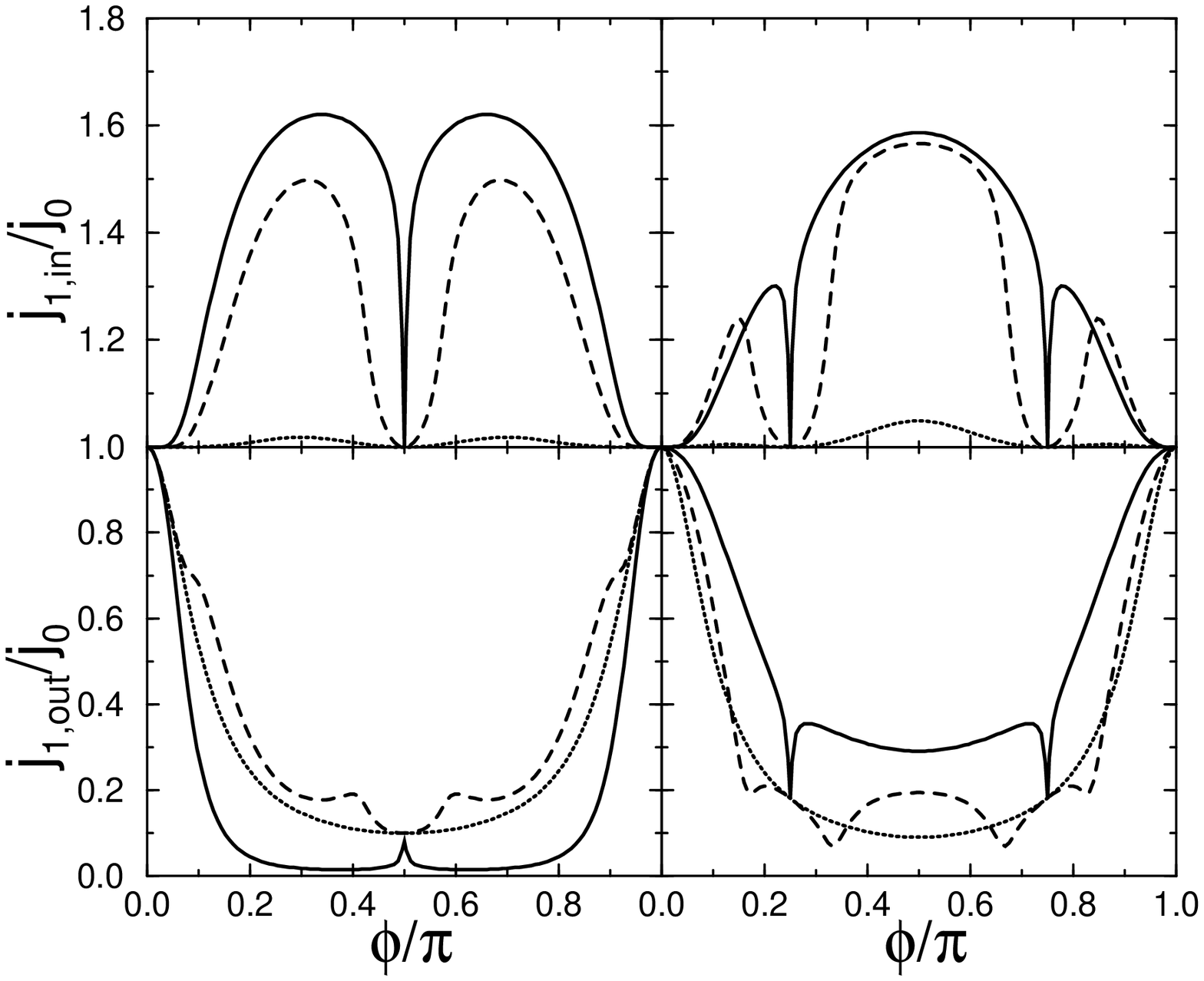}
}
\end{minipage}\\
\begin{minipage}{16.2cm}
\caption{ \label{A_sin_D=0.5_0.1}
Current densities in the injected beam (top panels) and in
the regularly reflected beam (bottom panels),
as a function of the impact angle for 3 energies: $\epsilon_b=0$ (full line),
$0.4\Delta_{max}$ (dashed) and $1.6 \Delta_{max}$ (dotted).
The left part of each picture is for a
(110) interface, and the right part of each picture for a (100) interface.
The left picture is for a transmission coefficient ${\cal D}_0=0.5$,
and the right picture for a transmission coefficient ${\cal D}_0=0.9$.
The temperature is $T=0.3T_c$,
and the mean free path $\ell =10 \xi_0 $.
}
\end{minipage}
\end{minipage}
\end{figure}
\rrule
\noindent
mean free path of $\ell =10 \xi_0$. 
The temperature was chosen 
$T=0.3T_c$, leading to a maximal gap of
$\Delta_{max}=2.29 T_c$. 
For the (100) orientation of the interface the order parameter is constant
in the superconductor for 
zero transmission and is suppressed at the interface for finite transmission.
In contrast for the (110) orientation the order parameter is 
suppressed to zero at the surface
for ${\cal D}=0$ and is suppressed to a finite value if 
${\cal D}$ is nonzero.\cite{bruder90} 
In the (100) orientation there is no  subgap
resonance, whereas a zero energy resonance typical for $d$-wave pairing at
properly oriented surfaces is present at (110) orientation.\cite{hu94}
Above the maximal gap the density of states is enhanced for (100) orientation. 
There is no such enhancement in the density of states at
the interface above the gap for (110) orientation.

\begin{figure}[h]
\begin{minipage}{16.3cm}
\begin{minipage}{8.1cm}
\centerline{
\epsfxsize8.0cm
\epsffile{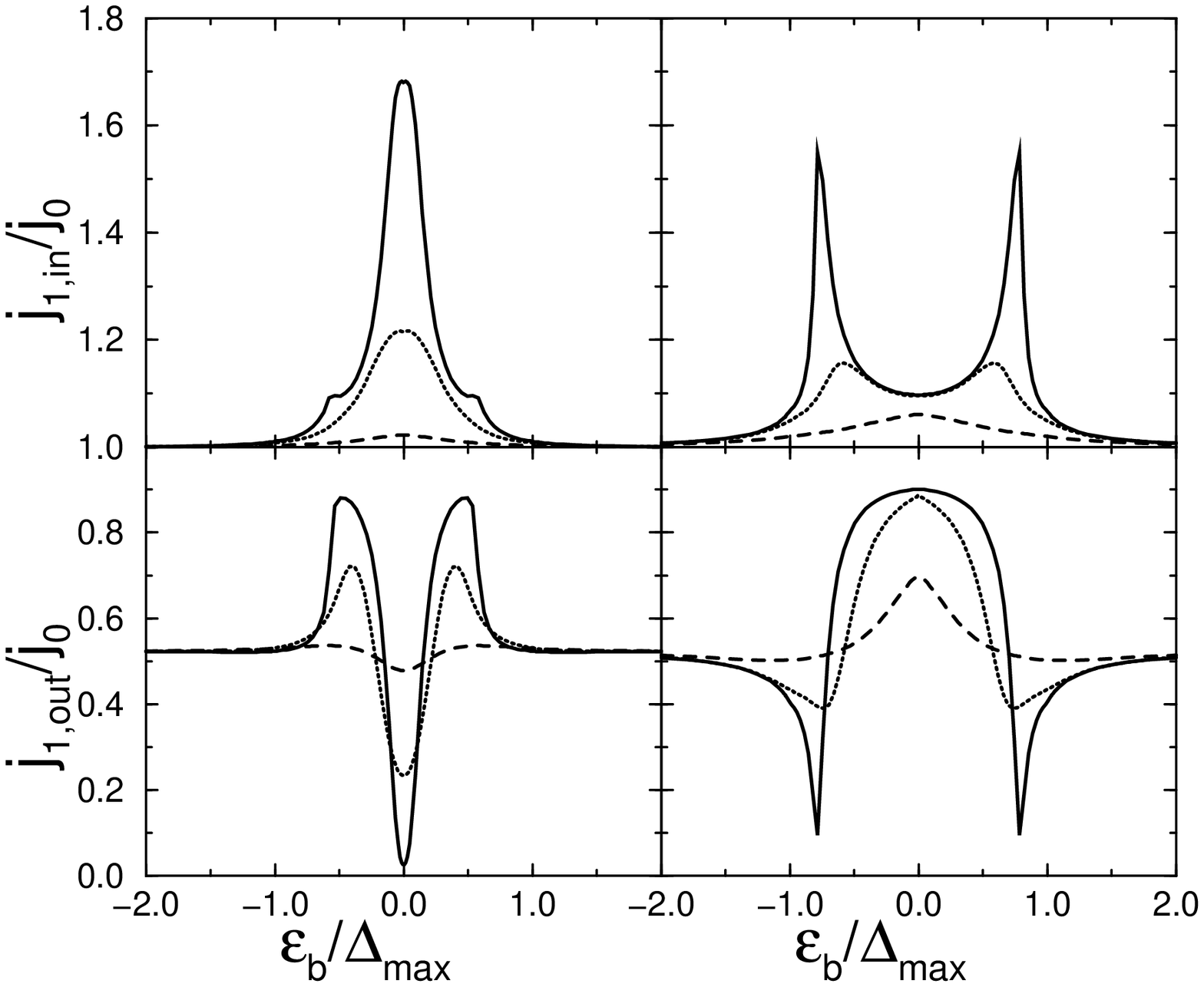}
}
\end{minipage}
\begin{minipage}{8.1cm}
\centerline{
\epsfxsize8.0cm
\epsffile{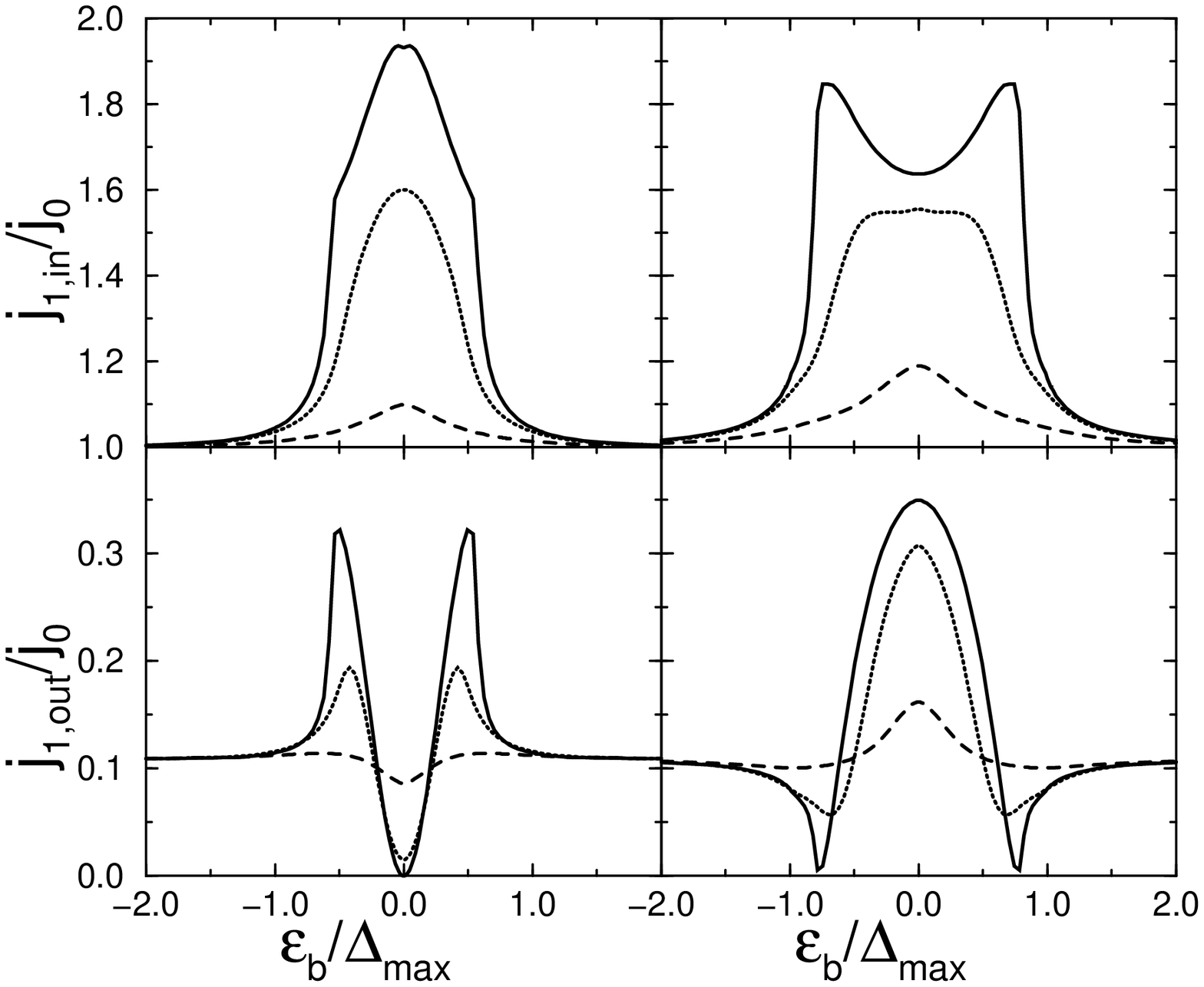}
}
\end{minipage}\\
\begin{minipage}{16.2cm}
\caption{ \label{B_sc_ofeps_D=0.9_T=0.3}
Current densities in the injected beam (top panels) and in
the regularly reflected beam (bottom panels),
as a function of energy for 3 different mean free path values for the
superconductor:
$\ell=100 \xi_0$ (full line), $10 \xi_0$ (dotted) and $2 \xi_0$ (dashed).
The impact angle is $\phi /\pi =0.4 $.
The left part of each picture is for a
(110) interface, and the right part of each picture for a (100) interface.
The left picture is for a transmission coefficient ${\cal D}_0=0.5$,
and the right picture for a transmission coefficient ${\cal D}_0=0.9$.
The temperature is $T=0.3T_c$. The values for the maximal gaps at this 
temperature are
$\Delta_{max}(\ell=100\xi_0)=2.13 T_c(\ell=100\xi_0)$,
$\Delta_{max}(\ell=10\xi_0)=2.29 T_c(\ell=10\xi_0)$,
$\Delta_{max}(\ell=2\xi_0)=2.85 T_c(\ell=2\xi_0)$.
}
\end{minipage}
\end{minipage}
\end{figure}

Figs. \ref{A_sin_D=0.5_0.1} and \ref{B_sc_ofeps_D=0.9_T=0.3}
show selected results of our calculations of Andreev
reflection at a contact between a normal metal and a $d$-wave
superconductor.  Our calculations are done  
for $T=0.3 T_c$, for three mean free
paths, $\ell =2\xi_0, 10 \xi_0,\ 100\xi_0 $, and for two
orientations of the interface. 
Fig. \ref{A_sin_D=0.5_0.1} shows for three  energies the
dependence of the excess current due to Andreev reflection (top panels)
and the regularly reflected current (bottom panels)
on the impact angle
for transmissions ${\cal D}_0=0.5 $ (left picture)
and for transmission ${\cal D}_0=0.9$ (right picture).
The positions of the gap nodes show up clearly in the Andreev reflection
amplitude, which breaks down for quasiparticles transmitted into the
nodal directions. The regular reflection approaches for the nodal direction
the value ${\cal R}(\phi)$. The width of this breakdown regions broadens 
with energy.
At energies above the maximal gap, $\Delta_{max}$, the Andreev amplitude
approaches zero and the regularly reflected amplitude approaches
the value ${\cal R}(\phi )$.
The dependence on the energy of the incoming quasiparticles
is shown for one representative impact angle in Fig. 
\ref{B_sc_ofeps_D=0.9_T=0.3} for three values of mean free path, again
for transmission ${\cal D}_0=0.5 $ (left) and for transmission ${\cal D}_0=0.9$ (right).
For the (100) interface 
as shown in Fig. \ref{A_sin_D=0.5_0.1} and \ref{B_sc_ofeps_D=0.9_T=0.3}, 
the behavior at low energies is clearly distinct from the 
behavior for a (110) interface. 
Whereas for a (110) interface the regular reflection is suppressed
for low energies, it is enhanced for a (100) interface.
The excess current shows a peak at low energies for the (110) interface,
but the (100) interface shows a minimum. The features at the gap edges are
small for the (110) orientation, but are strong for the (100) orientation.
And finally, the signal above the gap edges is small for a (110) interface
but extends well up to twice the gap for a (100) interface.

In the clean limit the
zero energy current density of the incoming beam is $j_{1,in}/j_0=
2$ for a (110) interface, and 
$j_{1,in}/j_0=2(1+{\cal R}^2)/(1+{\cal R})^2\le 2$
for a (100) interface;  
for the regularly reflected beam the zero
energy limit is $j_{1,out}/j_0=0$ for a (110) interface, and
$j_{1,out}/j_0= 4{\cal R}/(1+{\cal R})^2\ge {\cal R}$ for a (100) interface.
The values for the (100) interface coincide 
with the values for a
conventional isotropic $s$-wave superconductor, and are in agreement with
Blonder {\it et al.}\cite{blonder82} and Shelankov.\cite{shelankov84}
\lrule
\rrule
\noindent
Explicit values for the zero energy limits at a (100) 
surface are for perpendicular impact
$j_{1,in}/j_0=1.11$, $j_{1,out}/j_0=0.89$ for ${\cal D}=0.5 $, 
and $j_{1,in}/j_0=1.67$, $j_{1,out}/j_0=0.33 $ for ${\cal D}=0.9$. These 
values agree with our numerical calculations for mean 
free paths $\ell \ge 100 \xi_0$.  In contrast, the 
zero energy values for the (110) interface of two for the 
incoming and of zero 
for the reflected beam are very sensitive to impurity 
scattering. In fact, as can be seen from 
Fig.  \ref{B_sc_ofeps_D=0.9_T=0.3}, is the first value reduced to
about 1.2 for half transmission and a realistic mean free path of
ten coherence lengths, and the second value is larger than 0.2 in this case.
Also the structures around the gap edges for the (100) surface are
very sensitive to impurity scattering. For a mean free path of two
coherence lengths the Andreev signal is already strongly reduced, as our
calculations in Fig. \ref{B_sc_ofeps_D=0.9_T=0.3} show. This may
explain the small signal of only a few percent in many Andreev experiments.
The different behavior at low energies for the regular reflection
is the only remaining difference between (100) and (110) orientation for
mean free paths comparable to the coherence length for unconventional 
superconductors.
The suppression of the regularly 
reflected beam at low energies for all angles (except in nodal direction),
as seen for a (110) interface in the lower left panels of Figs.
\ref{A_sin_D=0.5_0.1} and \ref{B_sc_ofeps_D=0.9_T=0.3},
is a direct
consequence of the sign change of the order parameter during reflection
of quasiparticles. 
The origin of this effect is the same as for the zero energy resonance
(and follows from the Atiyah-Patodi-Singer theorem\cite{atiyah75}).
Both effects are destroyed by time reversal symmetry breaking and
both effects are washed out by impurity scattering.
However,
in contrast to the zero energy resonance, which is not an exact bound state
anymore for finite transmission even for zero impurity scattering, 
the strong 
suppression at low energies of the regularly reflected beam remains a stable
phenomenon for all transmissions in the clean limit. 
The effect is reduced by finite impurity scattering, and in this case
it is further reduced if the
transmission is comparable or smaller than the scattering rate.
Thus, the zero-energy resonance and the blocking of the regular reflection 
are two complementary phenomena: the first one is well established
only for interfaces with small transmissions, 
whereas the latter one is well established
at interfaces where the transmission is not too small.

The low-energy  behavior of the  regularly reflected beam can be used
to prove a sing change of the order parameter during reflection of the
quasiparticles at an interface. 
Specifically, our results show that for all impact angles
this reflection amplitude is always {\it above }
the normal state reflection, ${\cal R}(\phi)$, whereas 
for the (110) interface it is for all directions 
clearly {below} ${\cal R}(\phi)$ (the normal state reflection can be
obtained for a beam with $\epsilon_b$ well above the maximal gap).

\begin{figure}[h]
\begin{minipage}{16.3cm}
\begin{minipage}{8.1cm}
\centerline{
\epsfxsize8.0cm
\epsffile{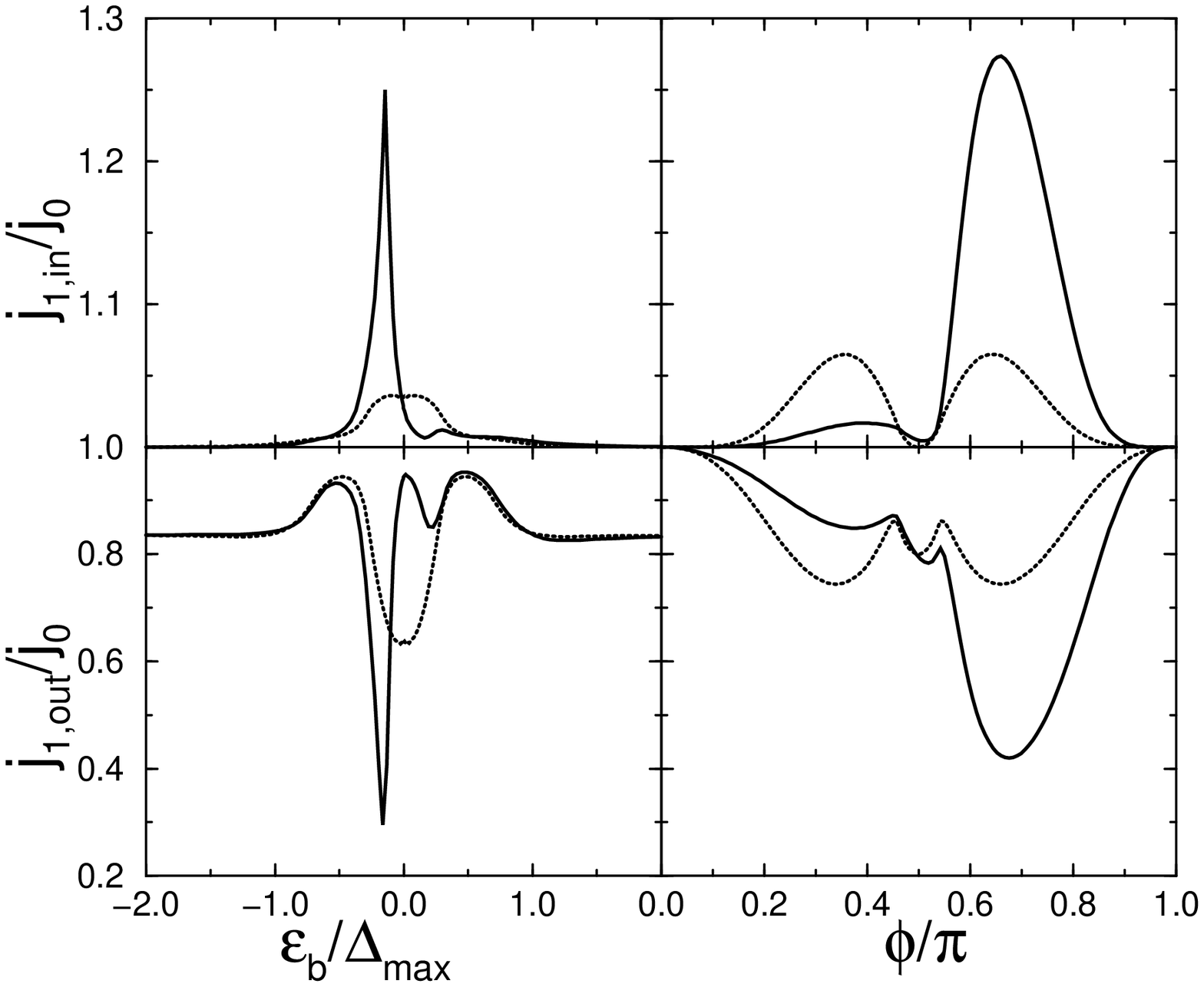}
}
\end{minipage}
\begin{minipage}{8.1cm}
\centerline{
\epsfxsize8.0cm
\epsffile{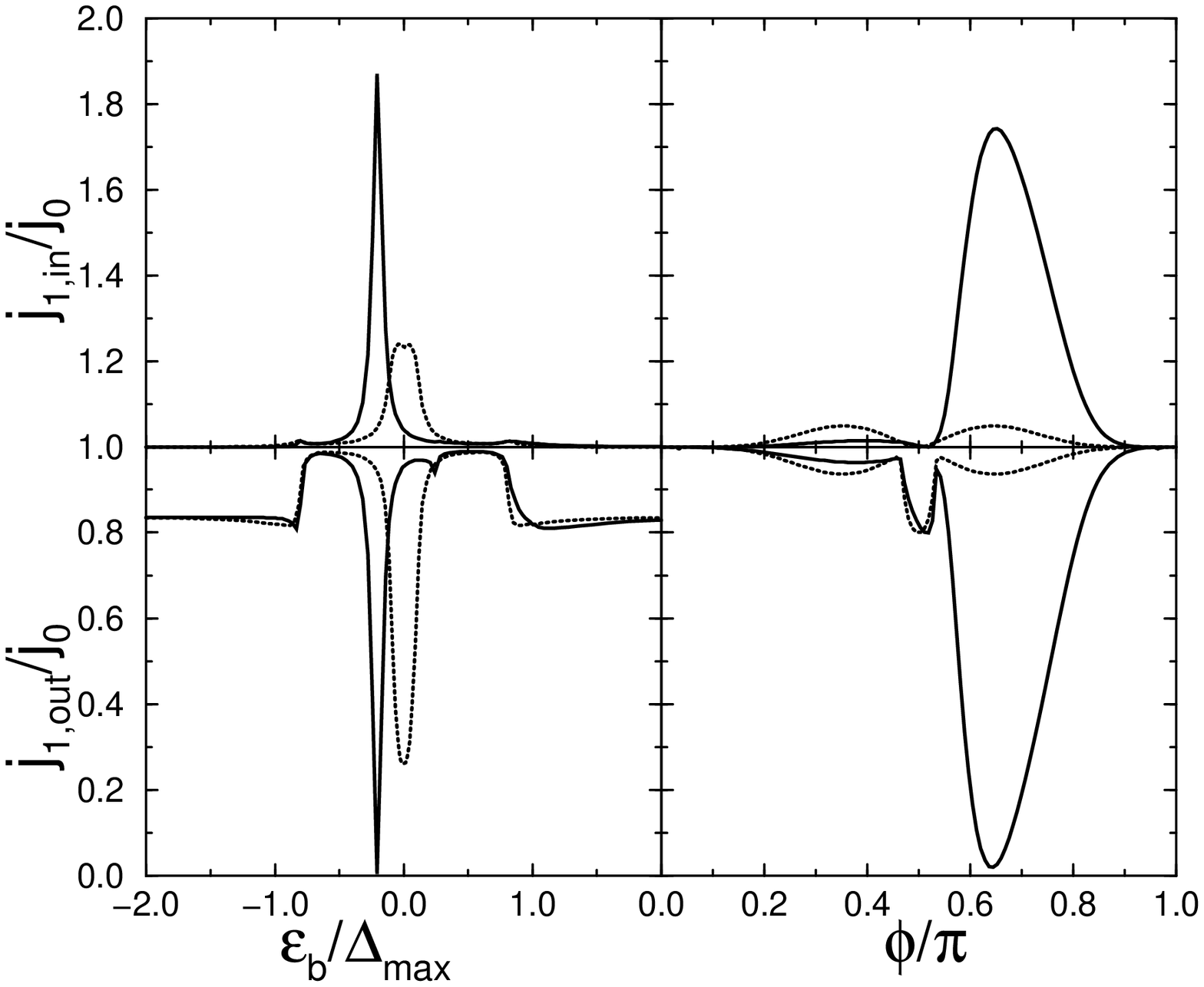}
}
\end{minipage}\\
\begin{minipage}{16.2cm}
\caption{ \label{B_dis_D=0.2_T=0.1_Ts=0.3_0.01}
Current densities in the injected beam (top panels) and in the
regularly reflected beam (bottom panels) at a (110) interface.
Dotted lines are for a temperature $T=0.3 T_d$, which is
above the transition temperature from
a $d$ to a $d+is$ state. Full lines are for a temperature $T=0.1 T_d$,
which is below this transition to the spontaneously
time reversal symmetry broken state.
The left picture is for mean free path $\ell =10 \xi_0 $, and
the right picture for mean free path $\ell =100 \xi_0 $.
The left part in each picture shows the energy dependence for
impact angle $\phi/\pi=0.348$, and the right part of each picture
shows the dependence on impact angle for $\epsilon = 0.2 \Delta_{max}$.
The transmission coefficient is ${\cal D}_0=0.2$. 
The subdominant transition temperature is $T_{s,0}=0.3 T_c $.
}
\end{minipage}
\end{minipage}
\end{figure}

Finally, we show that the low-energy 
suppression of the regular reflection and
enhancement of the excess current is a sensitive
test for time reversal symmetry breaking states.
In Fig. \ref{B_dis_D=0.2_T=0.1_Ts=0.3_0.01} we show our results
for a dominant $d$-wave coupled to a 
subdominant $s$-wave component.
\lrule
\rrule
\noindent
Below the interface transition\cite{matsumoto95,sigrist95,buchholtz95}
they couple to the spontaneously time reversal symmetry 
breaking state $d+is$,\cite{matsumoto95,sigrist95} where the $s$-wave component
is localized in a layer of a few coherence lengths near the interface. 
The left picture is for $\ell =10 \xi_0$ and 
the right picture for $\ell =100 \xi_0$. The coupling strength of the 
subdominant component is characterized by its `bare' 
transition temperature, $T_{s,0}$ (in the absence of the 
dominant component). 
This transition temperature
is reduced in the presence of the dominant component.\cite{matsumoto95,sigrist95,buchholtz95}
We chose $T_{s,0}=0.3T_d$, where $T_d$ is 
the transition temperature from the normal state to the pure $d$-wave state.
This value of $T_{s,0}$ 
is below the critical value for a possible transition into a
bulk $d+is$ state.\cite{matsumoto95} 
Nevertheless, the transition into a $d+is$ state localized near the
interface is possible.
According to our calculations
the subdominant component is strongly suppressed by 
finite transmission, so we chose a small value, ${\cal D}_0=0.2$.
The dotted curves show the current densities of the reflected beams
for a temperature $T=0.3 T_d$. 
The system is at this temperature
above the transition into the time reversal symmetry breaking state.
Full curves are for $T=0.1 T_d$, which corresponds to the 
interface $d+is$ state.
As can be seen from Fig. \ref{B_dis_D=0.2_T=0.1_Ts=0.3_0.01}, the
suppression of the reflection and the enhancement of Andreev reflection
are shifted to negative energies. Due to finite impurity scattering,
and resulting mixing of different momentum directions,
there is also a shadow-feature at positive energies. 
The broadening of the feature itself is reduced, leading to a much
sharper effect compared to the pure $d$-wave state.
The zero energy values for regular reflection
are changed in the $d+is$-state to almost one.
Also shown in Fig. \ref{B_dis_D=0.2_T=0.1_Ts=0.3_0.01} is the
dependence of the reflected current densities on the impact angle.
The small dip at $\phi=\pi/2$ is due to the fact that the energy of
the incoming particles is above the gap in these directions.
Below the transition into the $d+is$ state 
there is a strong anisotropy with respect to the interface normal.
This effect is a consequence of the spontaneous supercurrents at
the superconducting side of the interface.\cite{palumbo90} 
For an incoming particle beam
with a projection on the interface counter-moving with the current 
the regular reflection is strongly reduced compared to the
pure $d$-wave state, whereas the Andreev reflection is enhanced for
this case. For a beam with a projection co-moving with the
supercurrent these effects are absent or inverted. The
shift of the feature in the energy dependence of the
reflected current densities is determined by the Doppler shift of
the quasiparticle spectrum due to the spontaneous supercurrent
at the superconducting side of the interface. Thus, the {\it sign} of
the shift for a chosen impact angle
can be positive or negative dependent on the direction
of the spontaneous interface currents (similarly the asymmetry 
around the interface normal changes its sign).

\section{Conclusions}
\noindent
We have presented a new theoretical formulation of non-equilibrium
superconducting phenomena, including singlet and triplet pairing,
in terms of \state functions and distribution functions.
Our central results are equations (\ref{cgretav})-(\ref{keld2}),
together with boundary conditions at interfaces, equations
(\ref{bcc1})-(\ref{bck2}).
We used this formulation to present the theory for Andreev spectroscopy
at interfaces between a normal metal and an unconventional superconductor
in a transparent way. This formulation allows to include disorder
in a self-consistent manner. We proposed an anomalous suppression of
the regularly reflected quasiparticle beam as a test for 
time reversal symmetry breaking states. This test is especially
suitable for not too small transmission, where the 
zero energy interface resonances become ill-defined and cannot be used
as a test for time-reversal symmetry breaking anymore.

\acknowledgements
I gratefully acknowledge useful discussions with
L. Buchholtz, M. Fogelstr\"om, J. Moreno, D. Rainer, J. A. Sauls, and S.-K. Yip.
This work is supported by the National Science Foundation (DMR 91-20000)
through the Science and Technology Center for Superconductivity, and by the
Deutsche Forschungsgemeinschaft.
Part of the work was performed at the Aspen Center of Physics.
\lrule
\begin{appendix}
\rrule
\section{Notation}
\label{Notation}

The noncommutative $\qt $-product is defined in the following way
\beq
\label{timeconv}
\op{A} \qt \op{B}(\ep,t) =
\mbox{e}^{\frac{i\hbar }{2}(\partial_\ep^A \partial_t^B-\partial_t^A \partial_\ep^B)}
\op{A}(\ep,t) \op{B}(\ep,t) \quad .
\eeq
If one of the factors is both independent of $\ep $ and $t$, the
$\qt $-product reduces to the usual matrix product.\\
For Fourier transformed quantities ($t \to \om $) we have
\ber 
\op{A} \qt \op{B}(\ep,\om ) &=&
\int\limits_{-\infty}^{\infty} \frac{\mbox{d}\om'}{2\pi }
\frac{\mbox{d}\om''}{2\pi } \delta (\om' + \om'' - \om ) \times \nn \\
&&\times \op{A}(\epp{\ep}{\hbar \om'},\om'' ) \op{B}(\epm{\ep}{\hbar \om''},\om') \quad .
\eer
If $\op{A}(\ep,t)=\op{A}(\ep)$ is independent of $t$, that means,
$\op{A}$ is an equilibrium quantity, then
\beq
\op{A} \qt \op{B}(\ep,\om) =
\op{A}(\epp{\ep}{\hbar \om}) \op{B}(\ep,\om) \quad,
\eeq
and, analogously, if $\op{B}$ is an equilibrium quantity
\beq
\op{A} \qt \op{B}(\ep,\om) =
\op{A}(\ep,\om) \op{B}(\epm{\ep}{\hbar \om})\quad. 
\eeq
Also we generalize the commutator
\beq
[\op{A},\op{B}]_{\qt } = \op{A}\qt \op{B} - \op{B}\qt\op{A} \; .
\eeq
The Fermi surface average $\langle \cdots \rangle_{\vec{p}'}$
is defined by 
\beq
\langle \cdots \rangle_{\vec{p}'_f} = \frac{1}{N_f} \int 
\frac{d^2\vec{p}'_f }{(2\pi\hbar )^3 |\vec{v}_f(\vec{p}'_f)|} \cdots ,
\eeq
were $N_f$ is the total density of states at the Fermi surface in the
normal state,
\beq
N_f= \int \frac{d^2\vec{p}'_f }{(2\pi\hbar )^3 |\vec{v}_f(\vec{p}'_f)|},
\eeq
and $\vec{v}_f(\vec{p}'_f)$ is the normal state Fermi velocity at the position
$\vec{p}'_f$ on the Fermi surface
\beq
\vec{v}_f(\vec{p}'_f) = \frac{\partial \varepsilon (\vec{p})}{
\partial \vec{p}}\Bigg|_{\vec{p}=\vec{p}'_f}.
\eeq
Here, $ \varepsilon (\vec{p})$ describes the normal state dispersion of 
the quasiparticle band crossing the Fermi level at $\vec{p}'_f$.

\section{Projectors}\label{app_B}

Following Shelankov,\cite{shelankov85}
we introduce the following projectors
\ber
\label{proj}
\PPpm = \half \left( \ce \pm \frac{\; 1}{-i\pi} \; \cg \right) \quad .
\eer
From the normalization condition, 
$\cg \qt \cg =-\pi^2\, \ce $,
it follows that $\PPp $ and $\PPm $ are projection operators,
\ber
\label{projeig1}
\PPp \qt \PPp = \PPp \; , &\qquad& \PPm \qt \PPm = \PPm \; , 
\eer
and project orthogonal to each other
\ber
\label{projeig2}
\PPp + \PPm &=& \ce \quad , \\
\PPp \qt \PPm &=& \PPm \qt \PPp = \cz .
\eer
The quasiclassical Green's functions may be expressed in terms of $\PPp $ or $\PPm $,
\ber
\label{gqra1}
\cg &=& -i\pi \left( \PPp - \PPm \right) \nn \\
&=& -i\pi \left( 2\PPp - \ce \right)
= -i\pi \left( \ce - 2\PPm \right)
\, \, .
\eer
Equations of motion for the projectors can be extracted from the
corresponding equations for the quasiclassical Green's functions
\ber
\label{projbew}
\left[ \check{\ep } - \check{h},\PPpm \right]_{\qt } 
+i \hbar \qpartial \PPpm &=& \cz.
\eer
The Keldysh component of the Green's functions, $\gqk $, fulfills the
relation
$\gqr \qt \gqk + \gqk \qt \gqa = \op{0}$.
This implies $\Par \qt \gqk \qt \Paa = \op{0}$ and
$\Pbr \qt \gqk \qt \Pba = \op{0}$, leading to
\ber
\label{gqk1}
\gqk = \Par \qt \gqk \qt \Pba + \Pbr \qt \gqk \qt \Paa .
\eer
The value of $\Par \qt \gqk \qt \Pba$ does not determine $\gqk $
uniquely. It is possible to add 
$\Pbr \qt \op{A} + \op{B} \qt \Paa$ to $\gqk $ with any matrix function
$\op{A}$ and $\op{B} $ 
without changing $\Par \qt \gqk \qt \Pba$ (similarly
for $\Pbr \qt \gqk \qt \Paa $).
One can use this property to obtain a proper parameterization of $\gqk $
and to eliminate the unnecessary information in $\gqk$.
We write
\beq
\label{gqk2}
\gqk = -2\pi \; i \; \left( \Par \qt \XX \qt \Pba + \Pbr \qt \YY \qt \Paa \right)\; ,
\eeq
were $\XX$ and $\YY$ contain only one free spin matrix function as parameter.
The $\XX$ and $\YY$ 
have to fulfill fundamental symmetry relations, following from the
symmetry relations for the quasiclassical Green's function, $\check g$.
From the equations of motion of the Keldysh Green's functions
\ber
\label{gqkbew1}
(\ep \tc -\hr) \qt \gqk &-& \gqk \qt (\ep \tc - \ha ) 
+ i\hbar \qpartial \gqk = \nn \\
&-&(\gqr \qt \hk -
\hk \qt \gqa ) \; ,
\eer
we obtain the equations of motion for $\XX$ and $\YY$ using (\ref{projbew})
\lrule
\ber
\label{Xbew}
\Par \qt \bigg\{ (\ep \tc -\hr) \qt \XX - \XX \qt (\ep \tc - \ha ) +
\hk + i\hbar \qpartial \XX \bigg\} \qt \Pba &=& \op{0} \quad ,\\
\label{Ybew}
\Pbr \qt \bigg\{ (\ep \tc -\hr) \qt \YY - \YY \qt (\ep \tc - \ha ) -
\hk + i\hbar \qpartial \YY \bigg\} \qt \Paa &=& \op{0} \quad .
\eer
\rrule
Tracing these equations properly in the Nambu space
and respecting the symmetries of $\XX$ and $\YY$,
one obtains two equations for both undetermined 2x2
spin matrix functions, which parameterize $\XX$ and $\YY$.

Analogously we proceed for the linear response. 
From the $1^{st}$-order normalization
conditions (\ref{lrnorm}) we have
$\Para \qt
\dgqra \qt \Para=\op{0}$ 
and $\Pbra \qt \dgqra
\qt \Pbra=\op{0}$;
as a consequence
the spectral response, $\dgqra $,
can be written as 
\ber
&&\dgqra = \nonumber \\
\label{dXXYYra}
&&\mp 2\pi i \left[  \Para \qt \dXXra
\qt \Pbra- \Pbra \qt \dYYra
\qt \Para \right] \, .
\eer
Analogously, for the anomalous response the normalization condition
(\ref{lrnorm})
leads to
$\Par \qt \dgqan \qt \Paa =
\op{0}$ and $\Pbr \qt \dgqan
\qt \Pba =\op{0}$, so that 
$\dgqan $ can be written in the 
following form,
\begin{equation}
\label{dXXYYan}
\dgqan =-2\pi i \left[ \Par \qt
\dXXan \qt \Pba +\Pbr \qt
\dYYan \qt \Paa \right].
\end{equation}

\section{Parameter representations of the quasiclassical Green's 
functions}\label{app_C}

The projectors $\Par $ and $\Pbr $ may be parameterized in the following way
by complex spin matrices $\gar (\pf,\R,\ep,t)$ and $\gbr (\pf,\R,\ep,t)$
\ber
\label{par}
\Par &=& 
\matl \plus 1 \\ -\gbr \matlend 
\qt \left( 1-\gar \qt \gbr \right)^{-1} \qt
\mat 1, & \gar \matend ,\\
\label{pbr}
\Pbr &=& 
\matl -\gar \\ \plus 1 \matlend 
\qt \left( 1-\gbr \qt \gar \right)^{-1} \qt
\mat \gbr, & 1\matend .
\eer
Here $\left( 1+a \qt b \right)^{-1}$ is defined by
\beq
\label{id2}
\left( 1+a \qt b \right)^{-1} \qt \left( 1+a \qt b \right)=1
\, \, .
\eeq
One immediately proves $\Par \qt \Par = \Par$, $\Pbr \qt \Pbr =\Pbr $ and
$\Par \qt \Pbr =\Pbr \qt \Par =\op{0}$. A useful identity is
\beq
\label{id1}
\left( 1+ a \qt b \right)^{-1} \qt a = 
a \qt \left( 1+ b \qt a \right)^{-1} , 
\eeq
which may be used to obtain $\Par + \Pbr = 1$. The uniqueness of 
the projectors is ensured by the symmetry relations between the
matrix elements of the retarded and advanced Green's functions.
We may obtain the advanced Green's functions either by the
fundamental symmetry relation $ \gqa = \tc ( \gqr )^{\dagger} \tc$ or
analogously to the retarded case using advanced projectors
$\Paa = \tc ( \Pbr )^{\dagger} \tc$, $\Pba = 
\tc ( \Par )^{\dagger} \tc$
\ber
\label{paa}
\Paa &=& 
\matl - \gaa  \\ 1 \matlend 
\qt \left( 1-\gba \qt \gaa \right)^{-1} \qt
\mat  \gba, & 1 \matend ,\\
\label{pba}
\Pba &=& 
\matl 1 \\ - \gba  \matlend 
\qt \left( 1-\gaa \qt \gba \right)^{-1} \qt
\mat 1, & \plus \gaa \matend .
\eer
Here $\gaa=(\gbr)^{\dagger}$, $\gba = (\gar )^{\dagger}$ holds.
Introducing Eqs. (\ref{par}), (\ref{pbr}), (\ref{paa}), (\ref{pba})
into Eq. (\ref{projbew}),
and using $\ep \qt a + a \qt \ep = 2\ep a$ leads to the
transport equations for $\gara$ and $\gbra$,
Eqs. (\ref{cricc1}) and (\ref{cricc2}),
which are generalized Riccati differential equations. 
They are supplemented by properly chosen initial conditions.
The solutions $\gbra $, $\gara$ are introduced into Eqs.
(\ref{par}), (\ref{pbr}), (\ref{paa}), (\ref{pba}) to obtain the
quasiclassical Green's functions, Eqs. (\ref{cgretav}), (\ref{cnormfac} ), via
Eq. (\ref{gqra1}).

The solutions for the \state functions
in a homogeneous singlet superconductor
in equilibrium are,
\ber
\label{hom1}
\gara_{hom} = -\frac{\Dara \; \sign \epsilon }{|\Eara |+ \sqrt{(\Eara)^2+\Dara \Dbra }} \; , \\
\label{hom2}
\gbra_{hom} = \plus \frac{\Dbra \; \sign \epsilon }{|\Eara | + \sqrt{(\Eara)^2+\Dbra \Dara }} \; ,
\eer
where $\Eara = \epsilon - (\vara-\vbra)/2$. 
Note that $(\Dara \Dbra)$ is
proportional to the unit spin matrix and that in the clean limit
$(\Dara \Dbra) = -|\Delta|^2$.
In the presence of a constant
superflow with momentum $\vec{p}_s$ one has to make the replacement 
$\epsilon \to \epsilon - \vec{v}_f \cdot \vec{p}_s$.

Using this parameterization the following representation for the
Keldysh component with hermitian spin matrices $\xa(\pf,\R,\ep,t)$ and
$\xb(\pf,\R,\ep,t) $ is convenient.
Substituting
\beq
\label{kelgf1}
\XX = \mat \xa & 0 \\ 0 & 0 \matend , \quad 
\YY = \mat 0 & 0 \\ 0 & \xb \matend,
\eeq
into Eq. (\ref{gqk2}),
using the equation of motion for $\gqk $, Eq. (\ref{gqkbew1}),
leads to the transport equations for $\xa$ and $\xb $, Eqs. (\ref{keld1}) and
(\ref{keld2}).
Note that $\ep \qt a - a \qt \ep = i\hbar \partial_t a$.
These transport equations have to be supplemented by initial conditions.
For the Keldysh Green's function Eq. (\ref{kelgf1}) leads to Eq. (\ref{ckelgf2}).

It is possible to introduce Shelankov's
distribution functions\cite{shelankov85} $\Fa$ and $\Fb$,
which are given by the parameterization
\beq
\label{kelgf3}
\XX = \mat \Fa & 0 \\ 0 & \Fa \matend , \quad
\YY = \mat \Fb & 0 \\ 0 & \Fb \matend \; .
\eeq
They obey the symmetry relations
$\Fb(\pf,\R,\ep,t)=\Fa(-\pf,\R,-\ep,t)^{\ast }$, $\Fb(\pf,\R,\ep,t)
=\Fa(\pf,\R,\ep,t)^{\dagger }$.
The $\xa $ and $\xb $ are expressed in terms of them in the following way
\ber
\label{xaF}
\xa &=& \Big( \Fa - \gar \qt \Fa \qt \gba \Big) \; , \nonumber \\
\xb &=& \Big( \Fb - \gbr \qt \Fb \qt \gaa \Big) .
\eer
Using the introduced parameterization in terms of \state functions,
the transport equation for $\Fa$ has the form
\lrule
\ber
\label{keldFa}
(i\hbar \, \qpartial \Fa + i\hbar \, \partial_t \Fa ) &-&
\gar \qt (i\hbar \, \qpartial \Fa - i\hbar \, \partial_t \Fa ) \qt \gba + 
\nonumber \\[2mm]
&+& (-\var \qt \Fa +\Fa \qt \vaa + \vak) -\gar \qt (-\vbr \qt \Fa +\Fa \qt \vba- \vbk ) \qt \gba + \nonumber \\[2mm]
&-&\gar \qt (\Dbr \qt \Fa - \Fa \qt \Dba + \Dbk) + 
(\Dar \qt \Fa - \Fa \qt \Daa - \Dak) \qt \gba = 0 \quad .  
\eer
\rrule
The transport equation for 
$\Fb $ follows by application of the conjugation operation, Eq. (\ref{tildesymm}), to
this equation.
The $\gqk $ are obtained by introducing Eq. (\ref{xaF}) into Eq. (\ref{ckelgf2}).
The later parameterization is a convenient starting point for
perturbation theory from the equilibrium, because in the equilibrium
\beq
\qquad \Fa_{eq}=\tanh \frac{\ep}{2T}=-\Fb_{eq} 
\eeq
holds and all expression in the braces in Eq. (\ref{keldFa})
vanish independently.

Finally we make the connection to our parametrization in the
linear response. With the choises
\beq
\label{dXXra}
\dXXra = \mat 0 & \dgara \\ 0 & 0 \matend , \quad
\dYYra = \mat 0 & 0 \\ \dgbra & 0 \matend \; , 
\eeq
in Eq. (\ref{dXXYYra}), and
\beq
\label{dXXan}
\dXXan = \mat \dxaan & 0 \\ 0 & 0 \matend , \quad
\dYYan = \mat 0 & 0 \\ 0 & \dxban \matend \; , 
\eeq
in Eq. (\ref{dXXYYan}), we arrive at equations (\ref{xano})-(\ref{transpqcls4}).
With this parametrization the linear corrections to the
distibution spin matrices $\gara$, $\gbra $, $\xa $,
$\xb $ are given by $\dgara $, $\dgbra $, $\dxak $, and
$\dxbk $ respectively.

\section{Reflection and Transmission coefficents}
\label{refcoeff}
In the superconducting state
we define effective reflection and transmission coefficients by
\lrule
\ber
\label{BC_rho}
&&\ppar_{ij}=(1-\gar_i \qt \gbr_j) ,\qquad
\ppbr_{ij}=(1-\gbr_i \qt \gar_j) ,\qquad
(i,j=1,2),
\eer
\ber
\label{BC}
&&R_{1l}^{\ret}=
{\cal R} \ppar_{22} \qt
\Big( {\cal R}\ppar_{22}+{\cal D}\ppar_{12}\Big)^{-1} , \qquad
R_{1r}^{\ret}= 
\Big( {\cal R}\ppbr_{22}+{\cal D}\ppbr_{21}\Big)^{-1} \qt
{\cal R}\ppbr_{22} , \\
\label{BC1}
&&\tilde R_{1l}^{\ret}=
{\cal R} \ppbr_{22} \qt
\Big( {\cal R}\ppbr_{22}+{\cal D}\ppbr_{12}\Big)^{-1} , \qquad
\tilde R_{1r}^{\ret}= 
\Big( {\cal R}\ppar_{22}+{\cal D}\ppar_{21}\Big)^{-1} \qt
{\cal R}\ppar_{22} ,
\eer
\ber
\label{BC_A}
&&A_{1l}^{\ret}= 
(\gar_1-\gar_2) \qt
\Big( {\cal R}\ppbr_{22}+{\cal D}\ppbr_{21}\Big)^{-1} , \qquad
A_{1r}^{\ret}= 
\Big( {\cal R}\ppar_{22}+{\cal D}\ppar_{12}\Big)^{-1} \qt
(\gar_1-\gar_2) , \\
&&\tilde A_{1l}^{\ret}=
(\gbr_1-\gbr_2) \qt
\Big( {\cal R}\ppar_{22}+{\cal D}\ppar_{21}\Big)^{-1} , \qquad
\tilde A_{1r}^{\ret}= 
\Big( {\cal R} \ppbr_{22}+{\cal D}\ppbr_{12}\Big)^{-1} \qt
(\gbr_1-\gbr_2) .
\eer
\noindent
and $D_{1l}^{\ret}=1-R_{1l}^{\ret}$, $D_{1r}^{\ret}=1-R_{1r}^{\ret}$
$\tilde D_{1l}^{\ret}=1-\tilde R_{1l}^{\ret} $
$\tilde D_{1r}^{\ret}=1-\tilde R_{1r}^{\ret} $.
Analogously we define these quantities on the other side of the
interface by interchanging 1 and 2. 
Advanced quantities are given by
the same expressions with the change in the superscript $R\to A$.
The following relations are shown to hold
\ber
A_{1l}^{\ret}&= & \frac{R_{1l}^{\ret}}{\cal R} \qt \gar_1 - \frac{D_{1l}^{\ret}}{\cal D}
\qt \gar_2 
\quad
A_{1r}^{\ret}= \gar_1 \qt \frac{R_{1r}^{\ret}}{\cal R}
-\gar_2 \qt \frac{D_{1r}^{\ret}}{\cal D}
\; , \\
\tilde A_{1l}^{\ret}&= & \frac{\tilde R_{1l}^{\ret}}{\cal R} \qt \gbr_1 - \frac{\tilde D_{1l}^{\ret}}{\cal D}
\qt \gbr_2 
\quad
\tilde A_{1r}^{\ret}= \gbr_1 \qt \frac{\tilde R_{1r}^{\ret}}{\cal R}
-\gbr_2 \qt \frac{\tilde D_{1r}^{\ret}}{\cal D} \quad .
\eer
\rrule
As an example we consider the equilibrium spin-singlet case.
In equilibrium the $\qt $-product reduces to a matrix product.
In this case we can write 
$\gar = i \sigma_y \ga $, 
$\gbr = i \sigma_y \gb $, $\Gar = i\sigma_y \Ga $,
$\Gbr = i \sigma_y \Gb $,
where $\ga $, $\gb $, $\Ga $, $\Gb $ are scalar functions. 
The effective reflection and transmission coefficients are
\ber
R_1&=&{\cal R} \frac{1+\ga_2 \gb_2 }{1+{\cal R}\ga_2 \gb_2+{\cal D}\gb_2 \ga_1} \; , \\
D_1&=&{\cal D} \frac{1+\gb_2 \ga_1 }{1+{\cal R}\ga_2 \gb_2+{\cal D}\gb_2 \ga_1} \; , \\
\tilde R_1&=&{\cal R} \frac{1+\ga_2 \gb_2 }{1+{\cal R}\ga_2 \gb_2+{\cal D}\ga_2 \gb_1} \; ,  \\
\tilde D_1&=&{\cal D} \frac{1+\ga_2 \gb_1 }{1+{\cal R}\ga_2 \gb_2+{\cal D}\ga_2 \gb_1} \; , 
\eer
(and analogously for the other side of the interface), which 
fulfill $R_j+D_j=1$ and $\tilde R_j + \tilde D_j=1$ $(j=1,2)$.
\end{multicols}
\end{appendix}

\rrule

\end{multicols}


\begin{thebibliography}{10}
\bibitem{footn}
The second equalities in (\ref{bcc1}) and (\ref{bcc2})
can be verified using relations like
$({\cal D} \gar_1 + {\cal R}\gar_2) \qt (1-\gbr_2 \qt ({\cal D}\gar_1 + {\cal R} \gar_2))^{-1}=
(1-({\cal D}\gar_1 + {\cal R} \gar_2) \qt \gbr_2 )^{-1} \qt ({\cal D} \gar_1 + {\cal R}\gar_2)$.
\end{thebibliography}
\end{document}